\documentclass[12pt]{article}

\usepackage{epsf}
\usepackage{rotate}

\newcommand{\lsim}{
\mathrel{\hbox{\rlap{\hbox{\lower4pt\hbox{$\sim$}}}\hbox{$<$}}}}
\newcommand{\gsim}{
\mathrel{\hbox{\rlap{\hbox{\lower4pt\hbox{$\sim$}}}\hbox{$>$}}}}

\def\spose#1{\hbox to 0pt{#1\hss}}
\def\lsim{\mathrel{\spose{\lower 3pt\hbox{$\mathchar"218$}}
 \raise 2.0pt\hbox{$\mathchar"13C$}}}
\def\gsim{\mathrel{\spose{\lower 3pt\hbox{$\mathchar"218$}}
 \raise 2.0pt\hbox{$\mathchar"13E$}}}

\begin{document}

\begin{titlepage}

\begin{flushright}
CERN-TH/2003-084\\
hep-ph/0304027
\end{flushright}

\vspace{2cm}
\begin{center}
\boldmath
\large\bf New Strategies to Obtain Insights into CP Violation Through\\ 
\vspace{0.3truecm}
$B_s\to D_s^{\pm}K^\mp, D_s^{\ast\pm}K^\mp, ...$\ and 
$B_d\to D^{\pm}\pi^\mp, D^{\ast\pm}\pi^\mp, ...$\ Decays
\unboldmath
\end{center}

\vspace{1.2cm}
\begin{center}
Robert Fleischer\\[0.1cm]
{\sl Theory Division, CERN, CH-1211 Geneva 23, Switzerland}
\end{center}

\vspace{1.7cm}
\begin{abstract}
\vspace{0.2cm}\noindent
Decays of the kind $B_s\to D_s^{\pm}K^\mp, D_s^{\ast\pm}K^\mp, ...$\ and 
$B_d\to D^{\pm}\pi^\mp, D^{\ast\pm}\pi^\mp,...$\ allow us to probe $\phi_s+
\gamma$ and $\phi_d+\gamma$, respectively, involving the angle $\gamma$ of the 
unitarity triangle and the $B^0_q$--$\overline{B^0_q}$ mixing phases $\phi_q$ 
($q\in\{d,s\}$). Analysing these modes in a phase-convention-independent way, 
we find that their mixing-induced observables are affected by a subtle $(-1)^L$
factor, where $L$ denotes the angular momentum of the $B_q$ decay products, 
and derive bounds on $\phi_q+\gamma$. Moreover, we emphasize that ``untagged'' 
rates are an important ingredient for efficient determinations of weak phases, 
not only in the presence of a sizeable width difference $\Delta\Gamma_q$; 
should $\Delta\Gamma_s$ be sizeable, the combination of ``untagged'' with 
``tagged'' $B_s\to D_s^{\pm}K^\mp, D_s^{\ast\pm}K^\mp, ...$\ observables 
provides an elegant and {\it unambiguous} extraction of $\tan(\phi_s+\gamma)$, 
whereas the ``conventional'' determination of $\phi_s+\gamma$ is affected by
an eightfold discrete ambiguity. Finally, we propose a {\it combined} 
analysis of $B_s\to D_s^{\pm}K^\mp, D_s^{\ast\pm}K^\mp, ...$\ and 
$B_d\to D^{\pm}\pi^\mp,D^{\ast\pm}\pi^\mp,...$\ modes, which has important 
advantages, offering various interesting new strategies to extract $\gamma$ 
in an essentially unambiguous manner.
\end{abstract}

\vfill
\noindent
CERN-TH/2003-084\\
April 2003

\end{titlepage}

\thispagestyle{empty}
\vbox{}
\newpage
 
\setcounter{page}{1}

\section{Introduction}\label{sec:intro}
\setcounter{equation}{0}
The exploration of CP violation through studies of $B$-meson decays 
is one of the most exciting topics of present particle physics phenomenology, 
the main goal being to perform stringent tests of the Kobayashi--Maskawa
mechanism \cite{KM}. Here the central target is the unitarity triangle of 
the Cabibbo--Kobayashi--Maskawa (CKM) matrix, with its angles $\alpha$, 
$\beta$ and $\gamma$ (for a detailed review, see \cite{RF-PHYS-REP}).
Thanks to the efforts of the BaBar (SLAC) and Belle (KEK) collaborations, 
CP violation could recently be established in the neutral $B_d$-meson 
system with the help of $B_d\to J/\psi K_{\rm S}$ and similar decays 
\cite{CP-B-obs}. These modes allow us to determine $\sin\phi_d$, where the
present world average is given by $\sin\phi_d=0.734\pm0.054$ \cite{nir},
implying the twofold solution $\phi_d=(47^{+5}_{-4})^\circ \lor 
(133^{+4}_{-5})^\circ$ for the $B^0_d$--$\overline{B^0_d}$ mixing phase 
$\phi_d$, which equals $2\beta$ in the Standard Model. Here the former 
solution would be in perfect agreement with the ``indirect'' range 
following from the Standard-Model ``CKM fits'', 
$40^\circ\lsim\phi_d\lsim60^\circ$ \cite{CKM-fits}, whereas the 
latter would correspond to new physics \cite{FIM}. Measuring the sign of 
$\cos\phi_d$, the two solutions can be distinguished. Several strategies to 
accomplish this important task were proposed \cite{ambig}; an analysis 
using the time-dependent angular distribution of the decay products of 
$B_d\to J/\psi[\to\ell^+\ell^-] K^\ast[\to\pi^0K_{\rm S}]$ \cite{DDF,DFN} 
is already in progress at the $B$ factories \cite{itoh}.

\begin{figure}[h]
\begin{center}
\leavevmode
\epsfysize=4.4truecm 
\epsffile{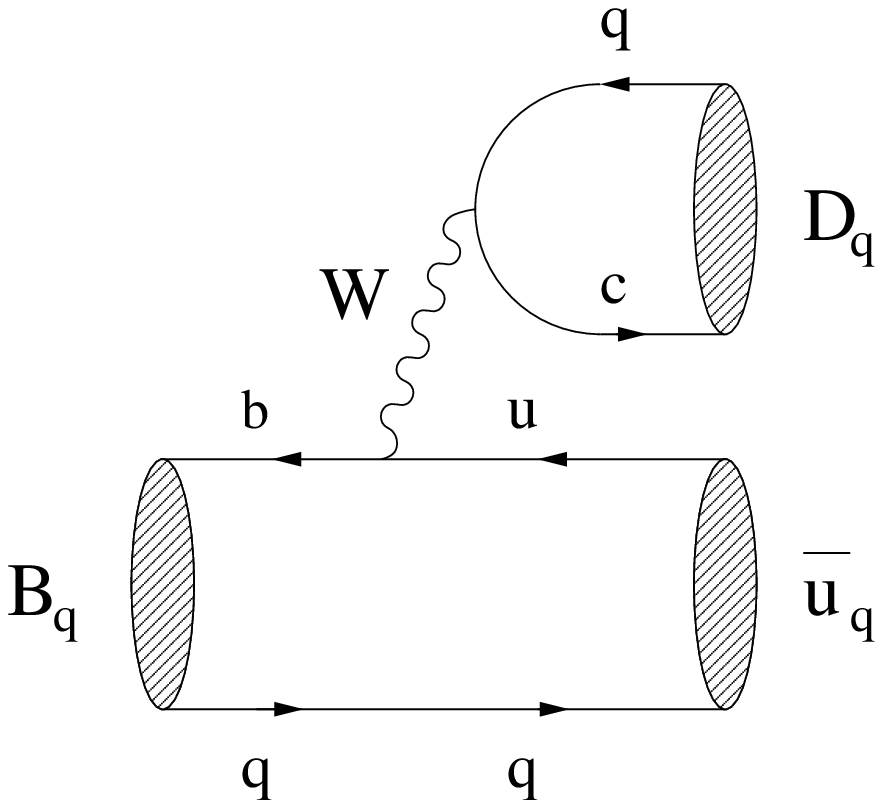} \hspace*{1truecm}
\epsfysize=4.4truecm 
\epsffile{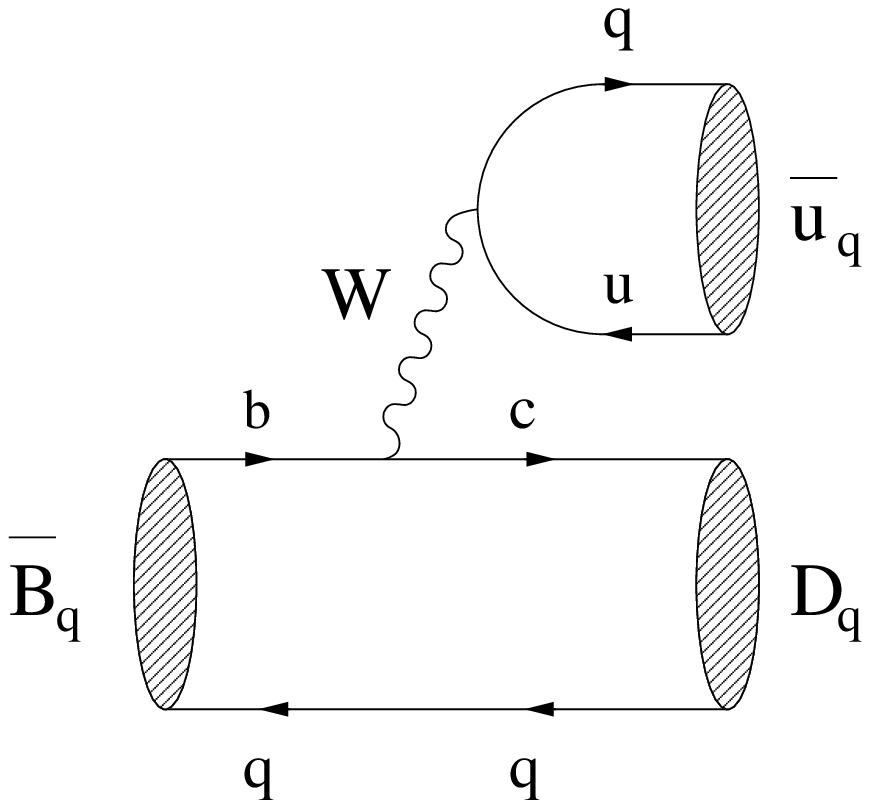}
\end{center}
\vspace*{-0.3truecm}
\caption{Feynman diagrams contributing to $B_q^0\to D_q \overline{u}_q$ and 
$\overline{B_q^0}\to D_q \overline{u}_q$.}\label{fig:BqDu}
\end{figure}

An important ingredient for the testing of the Kobayashi--Maskawa 
picture is provided by transitions of the kind $B_s\to D_s^{\pm}K^\mp, 
D_s^{\ast\pm}K^\mp, ...$\ \cite{BsDsK} and $B_d\to D^{\pm}\pi^\mp, 
D^{\ast\pm}\pi^\mp,...$\ \cite{BdDpi}, allowing theoretically clean
determinations of the weak phases $\phi_s+\gamma$ and $\phi_d+\gamma$, 
respectively, where $\phi_s$ is the $B_s$-meson counterpart of $\phi_d$, 
which is negligibly small in the Standard Model. It is convenient to 
write these decays generically as $B^0_q\to D_q\overline{u}_q$, so that
we may easily distinguish between the following cases: 
\begin{itemize}
\item $q=s$: $D_s\in\{D_s^+, D_s^{\ast+}, ...\}$, 
$u_s\in\{K^+, K^{\ast+}, ...\}$,
\item $q=d$: $D_d\in\{D^+, D^{\ast+}, ...\}$, 
$u_d\in\{\pi^+, \rho^+, ...\}$.
\end{itemize}
In the discussion given below, we shall only consider 
$B^0_q\to D_q\overline{u}_q$ decays, where at 
least one of the $D_q$, $\overline{u}_q$ states is a pseudoscalar meson. 
In the opposite case, for example the $B^0_s\to D_s^{\ast+}K^{\ast-}$ decay, 
the extraction of weak phases would require a complicated angular analysis 
\cite{FD2}--\cite{GPW}. If we look at Fig.~\ref{fig:BqDu}, we observe that 
$B^0_q\to D_q\overline{u}_q$ originates from colour-allowed tree-diagram-like 
topologies, and that also a $\overline{B^0_q}$ meson may decay into the same 
final state $D_q\overline{u}_q$. The latter feature leads to interference 
effects between $B^0_q$--$\overline{B^0_q}$ mixing and decay processes, 
allowing the extraction of $\phi_q+\gamma$ with an eightfold discrete 
ambiguity. Since $\phi_q$ can be straightforwardly fixed separately 
\cite{RF-PHYS-REP}, we may determine the angle $\gamma$ of the unitarity 
triangle from this CP-violating weak phase.

In Section~\ref{sec:ampl-obs}, we focus on the $B_q\to D_q\overline{u}_q$ 
decay amplitudes and rate asymmetries, and investigate the relevant hadronic 
parameters with the help of ``factorization''. In this section, we shall 
also point out that a subtle factor $(-1)^L$ arises in the expressions for 
the mixing-induced observables, where $L$ denotes the angular momentum of the 
$D_q\overline{u}_q$ system, and show explicitly the cancellation of 
phase-convention-dependent parameters within the factorization approach. 
After discussing the ``conventional'' extraction of 
$\phi_q+\gamma$ and the associated multiple discrete ambiguities 
in Section~\ref{sec:conv}, we emphasize the usefulness of ``untagged'' 
rate measurements for efficient determinations of weak phases from 
$B_q\to D_q\overline{u}_q$ decays in Section~\ref{sec:untagged}, and
suggest several novel strategies. In Section~\ref{sec:bounds}, we then derive 
bounds on $\phi_q+\gamma$, and illustrate their potential power with the 
help of a few numerical examples. In Section~\ref{sec:comb}, we propose a 
{\it combined} analysis of $B_{s}\to D_{s}\overline{u}_{s}$ 
and $B_{d}\to D_{d}\overline{u}_{d}$ modes, which has important advantages 
with respect to the conventional separate determinations of 
$\phi_s+\gamma$ and $\phi_d+\gamma$, offering various attractive new
avenues to extract $\gamma$ in an essentially unambiguous manner and
to obtain valuable insights into hadron dynamics. Finally, we conclude in 
Section~\ref{sec:concl}.

\section{Amplitudes, Rate Asymmetries and Factorization}\label{sec:ampl-obs}
\setcounter{equation}{0}
\subsection{Amplitudes}
The $B_q\to D_q\overline{u}_q$ decays are the colour-allowed counterparts
of the $B_s\to D\eta^{(')}, D\phi, ...$\ and $B_d\to D\pi^0, D\rho^0, ...$\ 
channels, which were recently analysed in detail in \cite{RF-BDf-1,RF-BDf-2}. 
If we follow the same avenue, and take also the Feynman diagrams shown in 
Fig.~\ref{fig:BqDu} into account, we may write
\begin{equation}\label{A-BbarDubar}
A(\overline{B^0_q}\to D_q\overline{u}_q)=\langle \overline{u}_qD_q|
{\cal H}_{\rm eff}(\overline{B^0_q}\to D_q\overline{u}_q)|
\overline{B^0_q}\rangle=\frac{G_{\rm F}}{\sqrt{2}}\overline{v}_q
\overline{M}_q,
\end{equation}
where the hadronic matrix element
\begin{equation}\label{Mbar-def}
\overline{M}_q\equiv 
\langle \overline{u}_qD_q|\overline{{\cal O}}_1^{\, q}\,{ C}_1(\mu)+
\overline{{\cal O}}_2^{\, q}\,{ C}_2(\mu)|\overline{B^0_q}\rangle
\end{equation}
involves the current--current operators
\begin{equation}
\overline{{\cal O}}_1^{\, q}\equiv
(\overline{q}_\alpha u_\beta)_{\mbox{{\scriptsize 
V--A}}}\left(\overline{c}_\beta b_\alpha\right)_{\mbox{{\scriptsize V--A}}},
\quad
\overline{{\cal O}}_2^{\, q}\equiv
(\overline{q}_\alpha u_\alpha)_{\mbox{{\scriptsize 
V--A}}}\left(\overline{c}_\beta b_\beta\right)_{\mbox{{\scriptsize V--A}}}.
\end{equation}
The CKM factors $\overline{v}_q$ are given by
\begin{equation}
\overline{v}_s\equiv V_{us}^\ast V_{cb}= A\lambda^3, \quad 
\overline{v}_d\equiv V_{ud}^\ast V_{cb}= A\lambda^2(1-\lambda^2/2),
\end{equation}
with (for the numerical value, see \cite{Andr-02})
\begin{equation}
A\equiv\frac{1}{\lambda^2}|V_{cb}|=0.83\pm0.02,
\end{equation}
and $\lambda\equiv |V_{us}|=0.22$ is the usual Wolfenstein parameter 
\cite{wolf}. 

On the other hand, the $B^0_q\to D_q\overline{u}_q$ decay amplitude
takes the following form:
\begin{eqnarray}
\lefteqn{A(B^0_q\to D_q\overline{u}_q)=\langle\overline{u}_qD_q|
{\cal H}_{\rm eff}(B^0_q\to D_q\overline{u}_q)|B^0_q\rangle}\nonumber\\
&&=\frac{G_{\rm F}}{\sqrt{2}}v_q^\ast \langle\overline{u}_qD_q|
{\cal O}_1^{q\dagger}\,{ C}_1(\mu)+{\cal O}_2^{q\dagger}\,{ C}_2(\mu)
|B^0_q\rangle,\label{ampl-BDubar1}
\end{eqnarray}
where we have to deal with the current--current operators 
\begin{equation}
{\cal O}_1^{q}\equiv(\overline{q}_\alpha c_\beta)_{\mbox{{\scriptsize V--A}}}
\left(\overline{u}_\beta b_\alpha\right)_{\mbox{{\scriptsize V--A}}},
\quad
{\cal O}_2^{q}\equiv(\overline{q}_\alpha c_\alpha)_{\mbox{{\scriptsize V--A}}}
\left(\overline{u}_\beta b_\beta\right)_{\mbox{{\scriptsize V--A}}},
\end{equation}
and the CKM factors $v_q$ are given by
\begin{equation}
v_s\equiv V_{cs}^\ast V_{ub}=A\lambda^3R_be^{-i\gamma}, 
\quad 
v_d\equiv V_{cd}^\ast V_{ub}=
-\left(\frac{A\lambda^4R_b}{1-\lambda^2/2}\right)e^{-i\gamma},
\end{equation}
with (for the numerical value, see \cite{Andr-02})
\begin{equation}\label{Rb-def}
R_b\equiv\left(1-\frac{\lambda^2}{2}\right)\frac{1}{\lambda}\left|
\frac{V_{ub}}{V_{cb}}\right|=\sqrt{\overline{\rho}^2+\overline{\eta}^2}
=0.39\pm0.04.
\end{equation}
If we introduce convention-dependent CP phases through
\begin{equation}\label{CP-phase-def}
({\cal CP})|F\rangle=e^{i\phi_{\rm CP}(F)}|\overline{F}\rangle,
\quad
({\cal CP})|\overline{F}\rangle=e^{-i\phi_{\rm CP}(F)}|F\rangle
\end{equation}
for $F\in\{B_q,D_q,u_q\}$, we obtain
\begin{equation}
({\cal CP})|D_q\overline{u}_q\rangle=(-1)^L e^{i[\phi_{\rm CP}(D_q)-
\phi_{\rm CP}(u_q)]}|\overline{D}_q u_q\rangle,
\end{equation}
where $L$ denotes the angular momentum of the $D_q\overline{u}_q$ state.
As we shall see below, the subtle $(-1)^L$ factor enters in mixing-induced
observables, and plays an important r\^ole for the extraction of weak
phases from these quantities in the presence of non-trivial angular momenta, 
for instance in the case of $B^0_d\to D^{\ast+}\pi^-$. In the literature, 
this factor does not show up explicitly in the context of 
$B^0_q\to D_q\overline{u}_q$ modes, but it was recently pointed out in the 
analysis of their colour-suppressed counterparts in 
\cite{RF-BDf-1,RF-BDf-2}. If we now employ, as in these
papers, the operator relations
\begin{equation}\label{CP-rel1}
({\cal CP})^\dagger ({\cal CP})=\hat 1,
\end{equation}
\begin{equation}
({\cal CP}){\cal O}_k^{q\dagger}({\cal CP})^\dagger=
{\cal O}_k^{q},
\end{equation}
we may rewrite (\ref{ampl-BDubar1}) as 
\begin{equation}\label{A-BDubar}
A(B^0_q\to D_q\overline{u}_q)=(-1)^L e^{i[\phi_{\rm CP}(B_q)-
\phi_{\rm CP}(D_q)+\phi_{\rm CP}(u_q)]}
\frac{G_{\rm F}}{\sqrt{2}}v_q^\ast M_q, 
\end{equation}
where
\begin{equation}\label{M-def}
M_{q}\equiv \langle u_q \overline{D}_q|{\cal O}_1^{q}\,{ C}_1(\mu)+
{\cal O}_2^{q}\,{ C}_2(\mu)|\overline{B^0_q}\rangle.
\end{equation}
It should be noted that also certain exchange topologies contribute to 
$B^0_q\to D_q\overline{u}_q$, $\overline{B^0_q}\to D_q\overline{u}_q$ 
transitions, which were -- for simplicity -- not shown in 
Fig.~\ref{fig:BqDu}. However, these additional diagrams do not affect the
phase structure of the amplitudes in (\ref{A-BbarDubar}) and 
(\ref{A-BDubar}), and manifest themselves only through tiny contributions
to the hadronic matrix elements $\overline{M}_q$ and $M_q$ given in
(\ref{Mbar-def}) and (\ref{M-def}), respectively. We shall come back to
these topologies in Subsection~\ref{subsec-DGam-0}, noting also how they
may be probed experimentally.

An analogous calculation for the $\overline{B^0_q}\to \overline{D}_qu_q$ 
and $B^0_q\to \overline{D}_qu_q$ processes yields
\begin{equation}\label{ampl-BbarDbaru}
A(\overline{B^0_q}\to \overline{D}_qu_q)=\frac{G_{\rm F}}{\sqrt{2}}v_q M_q
\end{equation}
\begin{equation}\label{ampl-BDu}
A(B^0_q\to \overline{D}_q u_q)=(-1)^L e^{i[\phi_{\rm CP}(B_q)+
\phi_{\rm CP}(D_q)-\phi_{\rm CP}(u_q)]}\frac{G_{\rm F}}{\sqrt{2}}
\overline{v}_q^\ast \overline{M}_q,
\end{equation}
where the same hadronic matrix elements as in the
$B^0_q\to D_q\overline{u}_q$, $\overline{B^0_q}\to D_q\overline{u}_q$
modes arise.

\subsection{Rate Asymmetries}
Let us first consider $B_q$ decays into $D_q\overline{u}_q$. Since both a 
$B^0_q$ and a $\overline{B^0_q}$ meson may decay into this state, we obtain 
a time-dependent rate asymmetry of the following form \cite{RF-PHYS-REP}:
\begin{eqnarray}
\lefteqn{\frac{\Gamma(B^0_q(t)\to D_q\overline{u}_q)-
\Gamma(\overline{B^0_q}(t)\to D_q\overline{u}_q)}{\Gamma(B^0_q(t)\to 
D_q\overline{u}_q)+\Gamma(\overline{B^0_q}(t)\to D_q\overline{u}_q)}}
\nonumber\\
&&=\left[\frac{C(B_q\to D_q\overline{u}_q)\cos(\Delta M_q t)+
S(B_q\to D_q\overline{u}_q)\sin(\Delta M_q t)}{\cosh(\Delta\Gamma_qt/2)-
{\cal A}_{\rm \Delta\Gamma}(B_q\to D_q\overline{u}_q)
\sinh(\Delta\Gamma_qt/2)}\right],\label{rate-asym}
\end{eqnarray}
where $\Delta M_q\equiv M_{\rm H}^{(q)}-M_{\rm L}^{(q)}>0$ is the mass
difference of the $B_q$ mass eigenstates $B_q^{\rm H}$ (``heavy'') and
$B_q^{\rm L}$ (``light''), and $\Delta\Gamma_q\equiv\Gamma_{\rm H}^{(q)}-
\Gamma_{\rm L}^{(q)}$ denotes their decay width difference, providing
the observable ${\cal A}_{\rm \Delta\Gamma}(B_q\to D_q\overline{u}_q)$. 
Before we turn to this quantity in the context of the ``untagged'' rates
discussed in Subsection~\ref{subsec:DGam}, let us first focus on
$C(B_q\to D_q\overline{u}_q)$ and $S(B_q\to D_q\overline{u}_q)$. These
observables are given by
\begin{equation}\label{Obs-expr}
C(B_q\to D_q\overline{u}_q)\equiv C_q=\frac{1-\bigl|\xi_q\bigr|^2}{1+
\bigl|\xi_q\bigr|^2}, \quad
S(B_q\to D_q\overline{u}_q)\equiv S_q=\frac{2\,\mbox{Im}\, \xi_q}{1+
\bigl|\xi_q\bigr|^2},
\end{equation}
where
\begin{equation}\label{xi-def}
\xi_q\equiv-e^{-i\phi_q}\left[e^{i\phi_{\rm CP}(B_q)}
\frac{A(\overline{B_q^0}\to D_q\overline{u}_q)}{A(B_q^0\to D_q\overline{u}_q)}
\right]
\end{equation}
measures the strength of the interference effects between the
$B^0_q$--$\overline{B^0_q}$ mixing and decay processes, involving
the CP-violating weak $B^0_q$--$\overline{B^0_q}$ mixing phase
\begin{equation}\label{phi-q-def}
\phi_q\equiv 2\,\mbox{arg}(V_{tq}^\ast V_{tb})
\stackrel{\rm SM}{=}\left\{\begin{array}{cl}
+2\beta={\cal O}(50^\circ) & ~~\mbox{($q=d$)}\\
-2\lambda^2\eta={\cal O}(-2^\circ) & 
~~\mbox{($q=s$).}
\end{array}\right.
\end{equation}
If we now insert (\ref{A-BbarDubar}) and (\ref{A-BDubar}) into (\ref{xi-def}), 
we observe that the convention-dependent phase $\phi_{\rm CP}(B_q)$ is 
cancelled through the amplitude ratio, and arrive at
\begin{equation}\label{xi-expr}
\xi_q=-(-1)^Le^{-i(\phi_q+\gamma)}\left[\frac{1}{x_qe^{i\delta_q}}\right],
\end{equation}
where
\begin{equation}\label{xq-def}
x_{s}\equiv R_b a_{s}, \quad 
x_{d}\equiv-\left(\frac{\lambda^2R_b}{1-\lambda^2}\right)a_{d},
\end{equation}
with
\begin{equation}\label{aq-def}
a_qe^{i\delta_q}\equiv e^{-i[\phi_{\rm CP}(D_q)-\phi_{\rm CP}(u_q)]}
\frac{M_q}{\overline{M}_q}.
\end{equation}
The convention-dependent phases $\phi_{\rm CP}(D_q)$ and $\phi_{\rm CP}(u_q)$
in (\ref{aq-def}) are cancelled through the ratio of hadronic matrix 
elements, so that $a_qe^{i\delta_q}$ is actually a physical observable.
Employing the factorization approach to deal with the hadronic matrix 
elements, we shall demonstrate this explicitly in 
Subsection~\ref{subsec:fact}. We may now apply (\ref{Obs-expr}), yielding
\begin{equation}
C_q=-\left[\frac{1-x_q^2}{1+x_q^2}\right], \quad
S_q=(-1)^L\left[\frac{2\,x_q\sin(\phi_q+\gamma+\delta_q)}{1+x_q^2}\right].
\end{equation} 

If we perform an analogous calculation for the decays into the CP-conjugate 
final state $\overline{D}_q u_q$, we obtain
\begin{equation}
\overline{\xi}_q=-e^{-i\phi_q}\left[e^{i\phi_{\rm CP}(B_q)}
\frac{A(\overline{B_q^0}\to \overline{D}_q u_q)}{A(B_q^0\to\overline{D}_qu_q)}
\right]=-(-1)^Le^{-i(\phi_q+\gamma)}\left[x_qe^{i\delta_q}\right],
\end{equation}
which implies
\begin{equation}
\overline{C}_q=+\left[\frac{1-x_q^2}{1+x_q^2}\right], \quad
\overline{S}_q=(-1)^L\left[\frac{2\,x_q
\sin(\phi_q+\gamma-\delta_q)}{1+x_q^2}\right],
\end{equation} 
where $\overline{C}_q\equiv C(B_q\to \overline{D}_q u_q)$ and
$\overline{S}_q\equiv S(B_q\to \overline{D}_q u_q)$. 

It should be noted that $\overline{\xi}_q$ and $\xi_q$ satisfy the relation
\begin{equation}\label{xi-rel}
\overline{\xi}_q\times\xi_q=e^{-i2(\phi_q+\gamma)},
\end{equation}
where the hadronic parameter $x_qe^{i\delta_q}$ cancels. Consequently, we
may extract $\phi_q+\gamma$ in a {\it theoretically clean} way from the 
corresponding observables. For our purposes, it will be convenient to 
introduce the following quantities:
\begin{equation}\label{Cp-def}
\langle C_q\rangle_+\equiv\frac{\overline{C}_q+C_q}{2}=0
\end{equation}
\begin{equation}\label{Cm-def}
 \langle C_q\rangle_-\equiv\frac{\overline{C}_q-C_q}{2}=\frac{1-x_q^2}{1+x_q^2}
\end{equation}
\begin{equation}\label{Sp-def}
\langle S_q\rangle_+\equiv\frac{\overline{S}_q+S_q}{2}=+(-1)^L
\left[\frac{2\,x_q\cos\delta_q}{1+x_q^2}\right]\sin(\phi_q+\gamma)
\end{equation}
\begin{equation}\label{Sm-def}
\langle S_q\rangle_-\equiv\frac{\overline{S}_q-S_q}{2}=-(-1)^L
\left[\frac{2\,x_q\sin\delta_q}{1+x_q^2}\right]\cos(\phi_q+\gamma).
\end{equation}
We observe that the factor $(-1)^L$ is crucial for the correctness of the
sign of the mixing-induced observable combinations $\langle S_q\rangle_+$ 
and $\langle S_q\rangle_-$. In particular, if we fix the sign of 
$\cos\delta_q$ through factorization arguments, we may determine 
the sign of $\sin(\phi_q+\gamma)$ from the measured sign of 
$\langle S_q\rangle_+$, providing valuable information. If we consider, 
for example, $B_s\to D_s^{\ast\pm}K^\mp$ or $B_d\to D^{\ast\pm}\pi^\mp$ 
modes, we have $L=1$, and obtain a non-trivial factor of $(-1)^1=-1$. On the 
other hand, we have $(-1)^0=+1$ in the case of $B_s\to D_s^{\pm} K^\mp$ or 
$B_d\to D^{\pm}\pi^\mp$ channels. Let us next analyse the hadronic parameter 
$a_qe^{i\delta_q}$ with the help of the factorization approach.

\subsection{Factorization}\label{subsec:fact}
Because of ``colour-transparency'' arguments \cite{col-trans,NS}, the
factorization of the hadronic matrix elements of four-quark operators
into the product of hadronic matrix elements of two quark currents can be 
nicely motivated for the decay $\overline{B^0_q}\to D_q\overline{u}_q$,
involving the matrix element $\overline{M}_q$. Recently, this picture 
could be put on a much more solid theoretical basis \cite{fact}. On the 
other hand, these arguments do not apply to the $\overline{B^0_q}\to 
\overline{D}_q u_q$ channel entering $M_q$, since there the spectator 
quark $q$ ends up in the $u_q$ meson, which is not ``heavy'' 
(see Fig.~\ref{fig:BqDu}). In order to analyse the hadronic 
parameter $a_qe^{i\delta_q}$ introduced in (\ref{aq-def}), it is
nevertheless instructive to apply ``na\"\i ve'' factorization not only 
to (\ref{Mbar-def}), but also to (\ref{M-def}), yielding
\begin{equation}\label{MA-fact1}
\left.\overline{M}_q\right|_{\rm fact}=a_1
\langle\overline{u}_q|(\overline{q}_\alpha u_\alpha)_{\mbox{{\scriptsize 
V--A}}}|0\rangle\langle D_q|(\overline{c}_\beta b_\beta)_{\mbox{{\scriptsize 
V--A}}}|\overline{B^0_q}\rangle
\end{equation}
\begin{equation}\label{MA-fact2}
\left.M_q\right|_{\rm fact}=a_1
\langle\overline{D}_q|(\overline{q}_\alpha c_\alpha)_{\mbox{{\scriptsize 
V--A}}}|0\rangle\langle u_q|(\overline{u}_\beta b_\beta)_{\mbox{{\scriptsize 
V--A}}}|\overline{B^0_q}\rangle,
\end{equation}
where
\begin{equation}
a_1=\frac{C_1(\mu_{\rm F})}{N_{\rm C}}+C_2(\mu_{\rm F})\approx1
\end{equation}
is the well-known phenomenological colour factor for colour-allowed decays
\cite{NS}, with a factorization scale $\mu_{\rm F}$ and a number $N_{\rm C}$ 
of quark colours. 

To be specific, let us consider the decays 
$\overline{B_s^0}\to D_s^{(\ast)+}K^-$ and 
$\overline{B_d^0}\to D^{(\ast)+}\pi^-$, i.e.\ $u_s=K^+$, $u_d=\pi^+$
and $D_s=D_s^{(\ast)+}$, $D_d=D^{(\ast)+}$. Using (\ref{CP-phase-def}) and 
(\ref{CP-rel1}), as well as
\begin{equation}
({\cal CP})\left[\overline{q}\gamma^\mu(1-\gamma_5)u\right]
({\cal CP})^\dagger=-\left[\overline{u}\gamma_\mu(1-\gamma_5)q\right],
\end{equation}
we obtain
\begin{eqnarray}
\langle\overline{u}_q|(\overline{q}_\alpha u_\alpha)_{\mbox{{\scriptsize 
V--A}}}|0\rangle&=&-e^{i\phi_{\rm CP}(u_q)}\langle u_q|(\overline{u}_\alpha 
q_\alpha)_{\mbox{{\scriptsize V--A}}}|0\rangle\label{CP-P}\\
\langle\overline{D}_q|
(\overline{q}_\alpha c_\alpha)_{\mbox{{\scriptsize V--A}}}|0\rangle&=&
-e^{i\phi_{\rm CP}(D_q)}\langle D_q|(\overline{c}_\alpha 
q_\alpha)_{\mbox{{\scriptsize V--A}}}|0\rangle
\end{eqnarray}
for the pseudoscalar mesons, and
\begin{equation}\label{CP-V}
\langle\overline{D}_q|
(\overline{q}_\alpha c_\alpha)_{\mbox{{\scriptsize V--A}}}|0\rangle=
+e^{i\phi_{\rm CP}(D_q)}\langle D_q|(\overline{c}_\alpha 
q_\alpha)_{\mbox{{\scriptsize V--A}}}|0\rangle
\end{equation}
for the vector mesons $D_s=D_s^{\ast+}$ and $D_d=D^{\ast+}$. If we now
use these expressions in (\ref{MA-fact1}) and (\ref{MA-fact2}), we see 
explicitly that the phase-convention-dependent factor in (\ref{aq-def})
is cancelled through the ratio of hadronic matrix elements, thereby
yielding a {\it convention-independent} result. In the case of the decays 
$B_s\to D_s^\pm K^\mp$ and $B_d\to D^\pm \pi^\mp$, we obtain
\begin{equation}
\left.a_s e^{i\delta_s}\right|_{\rm fact}
=\frac{f_{D_s}F^{(0)}_{B_sK^\pm}(M_{D_s}^2)(M_{B_s}^2-M_{K^\pm}^2)}{f_{K^\pm}
F^{(0)}_{B_sD_s}(M_{K^\pm}^2)(M_{B_s}^2-M_{D_s}^2)}
\end{equation}
and
\begin{equation}
\left.a_d e^{i\delta_d}\right|_{\rm fact}
=\frac{f_{D_d}F^{(0)}_{B_d\pi^\pm}(M_{D_d}^2)(M_{B_d}^2-
M_{\pi^\pm}^2)}{f_{\pi^\pm} F^{(0)}_{B_dD_d}
(M_{\pi^\pm}^2)(M_{B_d}^2-M_{D_d}^2)},
\end{equation}
respectively. If we apply heavy-quark arguments to the  
$\overline{B^0_s}\to D_s^+K^-$ and $\overline{B^0_d}\to D^+\pi^-$ 
modes \cite{NS,neubert-PHYS-REP}, we arrive at
\begin{equation}\label{as-fact-HQ}
\left.a_s e^{i\delta_s}\right|_{\rm fact}=
\frac{2f_{D_s}F^{(0)}_{B_sK^\pm}(M_{D_s}^2)(M_{B_s}^2-M_{K^\pm}^2)
\sqrt{M_{B_s}M_{D_s}}}{f_{K^\pm}\xi_s(w_s)(M_{B_s}-M_{D_s})
[(M_{B_s}+M_{D_s})^2-M_{K^\pm}^2]}
\end{equation}
\begin{equation}\label{ad-fact-HQ}
\left.a_d e^{i\delta_d}\right|_{\rm fact}=
\frac{2f_{D_d}F^{(0)}_{B_d\pi^\pm}(M_{D_d}^2)(M_{B_d}^2-M_{\pi^\pm}^2)
\sqrt{M_{B_d}M_{D_d}}}{f_{\pi^\pm}\xi_d(w_d)(M_{B_d}-M_{D_d})
[(M_{B_d}+M_{D_d})^2-M_{\pi^\pm}^2]},
\end{equation}
where the $\xi_q(w_q)$ are the Isgur--Wise functions describing
$\overline{B^0_q}\to D_q$ transitions, and 
\begin{equation}
w_s=\frac{M_{B_s}^2+M_{D_s}^2-M_{K^\pm}^2}{2M_{B_s}M_{D_s}},
\quad 
w_d=\frac{M_{B_d}^2+M_{D_d}^2-M_{\pi^\pm}^2}{2M_{B_d}M_{D_d}}.
\end{equation}
In the case of $B_s\to D_s^{\ast\pm} K^\mp$
and $B_d\to D^{\ast\pm}\pi^\mp$, we obtain accordingly
\begin{equation}\label{as-ast-fact-HQ}
\left.a_{s\ast} e^{i\delta_{s\ast}}\right|_{\rm fact}
=-\frac{f_{D_s^\ast}F^{(1)}_{B_sK^\pm}(M_{D_s^\ast}^2)}{f_{K^\pm}
A^{(0)}_{B_sD_s^\ast}(M_{K^\pm}^2)}=-\frac{2f_{D_s^\ast}
F^{(1)}_{B_sK^\pm}(M_{D_s^\ast}^2)\sqrt{M_{B_s}
M_{D_s^\ast}}}{f_{K^\pm}\xi_s(w_s^\ast)(M_{B_s}+M_{D_s^\ast})}
\end{equation}
and
\begin{equation}\label{ad-ast-fact-HQ}
\left.a_{d\ast} e^{i\delta_{d\ast}}\right|_{\rm fact}
=-\frac{f_{D_d^\ast}F^{(1)}_{B_d\pi^\pm}(M_{D_d^\ast}^2)}{f_{\pi^\pm}
A^{(0)}_{B_dD_d^\ast}(M_{\pi^\pm}^2)}=-\frac{2f_{D_d^\ast}
F^{(1)}_{B_d\pi^\pm}(M_{D_d^\ast}^2)\sqrt{M_{B_d}
M_{D_d^\ast}}}{f_{\pi^\pm}\xi_d(w_d^\ast)(M_{B_d}+M_{D_d^\ast})},
\end{equation}
respectively, where we have taken the relative minus sign between
(\ref{CP-P}) and (\ref{CP-V}) into account, and
\begin{equation}
w_s^\ast=\frac{M_{B_s}^2+M_{D_s^\ast}^2-M_{K^\pm}^2}{2M_{B_s}M_{D_s^\ast}},
\quad 
w_d^\ast=\frac{M_{B_d}^2+M_{D_d^\ast}^2-M_{\pi^\pm}^2}{2M_{B_d}M_{D_d^\ast}}.
\end{equation}

An important result of this exercise is 
\begin{equation}\label{phase-fact}
\left.\delta_{q}\right|_{\rm fact}=0^\circ, \quad
\left.\delta_{q\ast}\right|_{\rm fact}=180^\circ.
\end{equation}
Since factorization is expected to work well for 
$\overline{B^0_q}\to D_q\overline{u}_q$, in contrast to 
$\overline{B^0_q}\to \overline{D}_q u_q$, (\ref{phase-fact}) may in
principle receive large corrections, yielding sizeable CP-conserving 
strong phases. However, we may argue that we still have 
\begin{equation}\label{sgn-cos-del}
\cos\delta_{q}>0, \quad \cos\delta_{q\ast}<0,
\end{equation}
in accordance with the factorization prediction. This valuable information 
allows us to fix the sign of $\sin(\phi_q+\gamma)$ from (\ref{Sp-def}), where
the $(-1)^L$ factor plays an important r\^ole, as we already noted:
it is $+1$ and $-1$ for $B_s\to D_s^{\pm} K^\mp$, $B_d\to D^{\pm}\pi^\mp$ 
and $B_s\to D_s^{\ast\pm} K^\mp$, $B_d\to D^{\ast\pm}\pi^\mp$, respectively. 
Moreover, it should not be forgotten in this context that $x_s$ is 
positive, whereas $x_d$ is {\it negative} because of a factor of $-1$ 
originating from the ratio of CKM factors $v_d/\overline{v}_d^\ast$ 
(see (\ref{xq-def})).

Using, for instance, the Bauer--Stech--Wirbel form factors \cite{BSW},
we obtain $a_d=0.8$ and $a_{d\ast}=1.0$; if we take also (\ref{Rb-def}) 
and (\ref{xq-def}) into account, these values can be converted into 
$x_{d(\ast)}={\cal O}(-0.02)$, whereas $x_{s(\ast)}={\cal O}(0.4)$. In 
Section~\ref{sec:comb}, we shall have a closer look at the 
flavour-symmetry-breaking effects, which arise in the ratios 
$a_s/a_d$ and $a_{s\ast}/a_{d\ast}$.

It is useful to briefly compare these results with the situation of the
colour-suppressed counterparts of the $B_q\to D_q\overline{u}_q$ decays, 
the $B_s\to D\eta^{(')}, D\phi, ...$\ and $B_d\to D\pi^0, D\rho^0, ...$\ 
modes discussed in \cite{RF-BDf-1,RF-BDf-2}. Here factorization may receive 
sizeable corrections for each of the $B^0_q\to D^0f_q$ and 
$\overline{B^0_q}\to D^0f_q$ amplitudes. However, the corresponding 
hadronic matrix elements are actually very similar to one another, so 
that the factorized matrix elements cancel in the counterpart of 
$a_qe^{i\delta_q}$. Consequently, the thus obtained information on
the sign of the cosine of the corresponding strong phase difference 
$\delta_{f_q}$ appears to be a bit more robust than (\ref{sgn-cos-del}).

\boldmath
\section{Conventional Extraction of $\phi_q+\gamma$}\label{sec:conv}
\unboldmath
\setcounter{equation}{0}
We are now well prepared to discuss the ``conventional'' extraction of 
the CP-violating phase $\phi_q+\gamma$ from $B_q\to D_q\overline{u}_q$ 
decays \cite{BsDsK,BdDpi}.
As we have already noted, because of (\ref{xi-rel}), it is obvious that
these modes and their CP conjugates provide a theoretically clean 
extraction of this phase. Using (\ref{Cm-def}), we may -- in 
principle -- determine $x_q$ through
\begin{equation}\label{xq-def-conv}
x_q=\eta_q\sqrt{\frac{1-\langle C_q\rangle_-}{1+\langle C_q\rangle_-}},
\end{equation}
where 
\begin{equation}
\eta_q=\left\{
\begin{array}{ll}
+1 & ~~\mbox{($q=s$)}\\
-1 & ~~\mbox{($q=d$)}
\end{array}
\right.
\end{equation}
takes into account the minus sign appearing in (\ref{xq-def}) for 
$q=d$. Using the knowledge of $x_q$, we may extract the following 
quantities from the combinations of the mixing-induced observables 
introduced in (\ref{Sp-def}) and (\ref{Sm-def}):
\begin{equation}\label{s-p}
s_+\equiv(-1)^L\left[\frac{1+x_q^2}{2\,x_q}\right]\langle S_q\rangle_+
=+\cos\delta_q\sin(\phi_q+\gamma)
\end{equation} 
\begin{equation}\label{s-m}
s_-\equiv(-1)^L\left[\frac{1+x_q^2}{2\,x_q}\right]\langle S_q\rangle_-
=-\sin\delta_q\cos(\phi_q+\gamma),
\end{equation} 
which allow us to determine $\sin^2(\phi_q+\gamma)$ with the help of 
\begin{equation}\label{sin2phi-det}
\sin^2(\phi_q+\gamma)=\frac{1}{2}\left[(1+s_+^2-s_-^2)\pm
\sqrt{(1+s_+^2-s_-^2)^2-4s_+^2}\right].
\end{equation}
This relation implies a fourfold solution for $\sin(\phi_q+\gamma)$.
Since each value of this quantity corresponds to a twofold solution for
$\phi_q+\gamma$, the extraction of this phase suffers, in general, from 
an eightfold discrete ambiguity. If we employ (\ref{sgn-cos-del}) and 
(\ref{s-p}), the measured sign of $s_+$ allows us to fix the sign of 
$\sin(\phi_q+\gamma)$, thereby reducing the discrete ambiguity for 
the value of $\phi_q+\gamma$ to a fourfold one. Needless to note that
these unpleasant ambiguities significantly reduce the power to search for 
possible signals of new physics.

Another disadvantage is that the determination of the hadronic parameter
$x_q$ through (\ref{xq-def-conv}) requires the experimental resolution of 
small $x_q^2$ terms in (\ref{Cm-def}). In the $q=s$ case, we na\"\i vely 
expect $x_s^2={\cal O}(0.16)$, so that this may actually be possible, though 
challenging.\footnote{Note that non-factorizable effects may well lead to a
significant reduction or enhancement of $x_s$.} On the other hand, it is 
practically impossible to resolve the $x_d^2={\cal O}(0.0004)$ terms, i.e.\
(\ref{Cm-def}) is not effective in the $q=d$ case. However, it may well be 
possible to measure the observable combinations $\langle S_d\rangle_+$ and 
$\langle S_d\rangle_-$, since these quantities are proportional to 
$x_d={\cal O}(-0.02)$. In this respect, $B_d\to D^{\ast\pm}\pi^\mp$ 
channels are particularly promising, since they exhibit large branching ratios 
at the $10^{-3}$ level and offer a good reconstruction of the 
$D^{\ast\pm}\pi^\mp$ states with a high efficiency and modest backgrounds 
\cite{LHC-Report,BABAR-Book}. In order to solve the problem of the extraction 
of $x_d$, which was also addressed in \cite{BdDpi}, we shall propose the use 
of ``untagged'' decay rates, where we do not distinguish between initially, 
i.e.\ at time $t=0$, present $B^0_d$ or $\overline{B^0_d}$ mesons. Also 
in the case of $q=s$, alternatives to (\ref{xq-def-conv}) for an efficient 
determination of $x_s$ are obviously desirable.

\section{Closer Look at ``Untagged'' Rates}\label{sec:untagged}
\setcounter{equation}{0}
\boldmath
\subsection{New Strategy Employing $\Delta\Gamma_q$}\label{subsec:DGam}
\unboldmath
As we have seen in (\ref{rate-asym}), the width difference 
$\Delta\Gamma_q$ of the $B_q$ mass eigenstates provides another 
observable, ${\cal A}_{\rm \Delta\Gamma}(B_q\to D_q\overline{u}_q)$, 
which is given by
\begin{equation}\label{ADGam}
{\cal A}_{\rm \Delta\Gamma}(B_q\to D_q\overline{u}_q)=
\frac{2\,\mbox{Re}\,\xi_q}{1+\bigl|\xi_q\bigr|^2}.
\end{equation}
This quantity is, however, not independent from $C(B_q\to D_q\overline{u}_q)$ 
and $S(B_q\to D_q\overline{u}_q)$, satisfying the relation
\begin{equation}\label{Obs-rel}
\Bigl[C(B_q\to D_q\overline{u}_q)\Bigr]^2+
\Bigl[S(B_q\to D_q\overline{u}_q)\Bigr]^2+
\Bigl[{\cal A}_{\Delta\Gamma}(B_q\to D_q\overline{u}_q)\Bigr]^2=1.
\end{equation}
Interestingly, ${\cal A}_{\rm \Delta\Gamma}(B_q\to D_q\overline{u}_q)$
could be determined from the ``untagged'' rate
\begin{eqnarray}
\lefteqn{\langle\Gamma(B_q(t)\to D_q\overline{u}_q)\rangle\equiv
\Gamma(B^0_q(t)\to D_q\overline{u}_q)+
\Gamma(\overline{B^0_q}(t)\to D_q\overline{u}_q)}\nonumber\\
&&=\left[\Gamma(B^0_q\to D_q\overline{u}_q)+
\Gamma(\overline{B^0_q}\to D_q\overline{u}_q)\right]\nonumber\\
&&\times\left[\cosh(\Delta\Gamma_qt/2)-{\cal A}_{\rm \Delta\Gamma}
(B_q\to D_q\overline{u}_q)\,\sinh(\Delta\Gamma_qt/2)\right]
e^{-\Gamma_qt},\label{untagged}
\end{eqnarray}
where the oscillatory $\cos(\Delta M_qt)$ and $\sin(\Delta M_qt)$ terms 
cancel, and $\Gamma_q\equiv(\Gamma_{\rm H}^{(q)}+\Gamma_{\rm L}^{(q)})/2$ 
denotes the average decay width \cite{dunietz}. In the case of the 
$B_d$-meson system, the width difference is negligibly small, so that the 
time evolution of (\ref{untagged}) is essentially given by the well-known 
exponential $e^{-\Gamma_dt}$. On the other hand, the width difference 
$\Delta \Gamma_s$ of the $B_s$-meson system may be as large as 
${\cal O}(-10\%)$ (for a recent review, see \cite{BeLe}), and 
may hence allow us to extract 
${\cal A}_{\rm \Delta\Gamma}(B_s\to D_s\overline{u}_s)$.

Inserting (\ref{xi-expr}) into (\ref{ADGam}), we obtain
\begin{equation}
{\cal A}_{\rm \Delta\Gamma}(B_s\to D_s\overline{u}_s)\equiv 
{\cal A}_{\rm \Delta\Gamma_s}=-(-1)^L
\left[\frac{2\,x_s\cos(\phi_s+\gamma+\delta_s)}{1+x_s^2}\right],
\end{equation}
and correspondingly
\begin{equation}
{\cal A}_{\rm \Delta\Gamma}(B_s\to \overline{D}_s u_s)\equiv 
\overline{{\cal A}}_{\rm \Delta\Gamma_s}=-(-1)^L
\left[\frac{2\,x_s\cos(\phi_s+\gamma-\delta_s)}{1+x_s^2}\right],
\end{equation}
which yields
\begin{equation}\label{def-Ap}
\langle{\cal A}_{\rm \Delta\Gamma_s}\rangle_+
\equiv\frac{\overline{{\cal A}}_{\rm \Delta\Gamma_s}+
{\cal A}_{\rm \Delta\Gamma_s}}{2}=
-(-1)^L\left[\frac{2\,x_s\cos\delta_s}{1+x_s^2}
\right]\cos(\phi_s+\gamma)
\end{equation}
\begin{equation}\label{def-Am}
\langle{\cal A}_{\rm \Delta\Gamma_s}\rangle_-\equiv
\frac{\overline{{\cal A}}_{\rm \Delta\Gamma_s}-
{\cal A}_{\rm \Delta\Gamma_s}}{2}=
-(-1)^L\left[\frac{2\,x_s\sin\delta_s}{1+x_s^2}
\right]\sin(\phi_s+\gamma).
\end{equation}
If we compare now (\ref{def-Ap}) and (\ref{def-Am}) with (\ref{Sp-def}) and 
(\ref{Sm-def}), respectively, we observe that the same hadronic factors 
enter in these mixing-induced observables, and obtain
\begin{equation}\label{extr-1}
\frac{\langle S_s\rangle_+}{\langle{\cal A}_{\rm \Delta\Gamma_s}\rangle_+}
=-\tan(\phi_s+\gamma)
\end{equation}
\begin{equation}\label{extr-2}
\frac{\langle{\cal A}_{\rm \Delta\Gamma_s}\rangle_-}{\langle S_s\rangle_-}
=+\tan(\phi_s+\gamma),
\end{equation}
implying the consistency relation
\begin{equation}
\langle{\cal A}_{\rm \Delta\Gamma_s}\rangle_+
\langle{\cal A}_{\rm \Delta\Gamma_s}\rangle_-
=-\langle S_s\rangle_+\langle S_s\rangle_-.
\end{equation}
Should $\delta_s$ take values around $0^\circ$ or $180^\circ$, as in
factorization (see (\ref{phase-fact})), we
may extract $\tan(\phi_s+\gamma)$ from
(\ref{extr-1}), whereas we could use (\ref{extr-2}) in the opposite case 
of $\delta_s$ being close to $+90^\circ$ or $-90^\circ$. The strong
phase itself can be determined from
\begin{equation}
\tan\delta_s=\frac{\langle S_s\rangle_-}{\langle{\cal A}_{\rm 
\Delta\Gamma_s}\rangle_+}=-\frac{\langle{\cal A}_{\rm 
\Delta\Gamma_s}\rangle_-}{\langle S_s\rangle_+}.
\end{equation}
The values of $\tan(\phi_s+\gamma)$ and $\tan\delta_s$ thus extracted imply
twofold solutions for $\phi_s+\gamma$ and $\delta_s$, respectively, which
should be compared with the eightfold solution for $\phi_s+\gamma$
following from (\ref{sin2phi-det}). Using (\ref{sgn-cos-del}), 
we may immediately fix $\delta_s$ unambiguously, and may determine 
the sign of $\sin(\phi_s+\gamma)$ with the help of the measured sign
of $\langle S_s\rangle_+$ from (\ref{Sp-def}), thereby resolving the 
twofold ambiguity for the value of $\phi_s+\gamma$. On the other hand, 
the ``conventional'' approach discussed in Section~\ref{sec:conv} would 
still leave a fourfold ambiguity for this phase, as we shall illustrate in
Section~\ref{sec:bounds}. Finally, we may of course also determine 
$x_s$ from one of the $\langle S_s\rangle_\pm$ or 
$\langle{\cal A}_{\rm \Delta\Gamma_s}\rangle_\pm$ observables. 

We observe that the combination of the ``tagged'' mixing-induced 
observables $\langle S_s\rangle_\pm$ with their ``untagged'' 
counterparts $\langle{\cal A}_{\rm \Delta\Gamma_s}\rangle_\pm$ provides
an elegant determination of $\phi_s+\gamma$ in an essentially unambiguous 
manner. In \cite{FD2}, strategies to determine this phase from untagged 
$B_s$ data samples only were proposed, which employ angular distributions 
of decays of the kind $B_s\to D_s^{\ast\pm} K^{\ast\mp}$ and are hence 
considerably more involved. Another important advantage of our new strategy 
is that both $\langle S_s\rangle_\pm$ and 
$\langle{\cal A}_{\rm \Delta\Gamma_s}\rangle_\pm$ are proportional to 
$x_s$. Consequently, the extraction of $\phi_s+\gamma$ does not require 
the resolution of $x_s^2$ terms.\footnote{A similar feature is 
also present in the ``untagged'' $B_s\to D_s^{\ast\pm} K^{\ast\mp}$ strategy 
proposed in \cite{FD2}, and in the ``tagged'' analysis in \cite{LSS}, 
employing the angular distribution of the $D_s^{\ast\pm}$, $K^{\ast\mp}$ 
decay products.} On the other hand, we have to rely on a sizeable width 
difference $\Delta\Gamma_s$, which may be too small to make an 
extraction of $\langle{\cal A}_{\rm \Delta\Gamma_s}\rangle_\pm$ experimentally 
feasible. In the presence of CP-violating new-physics contributions to 
$B^0_s$--$\overline{B^0_s}$ mixing, manifesting themselves through a sizeable 
value of $\phi_s$, $\Delta\Gamma_s$ would be further reduced, as follows 
\cite{grossman}:
\begin{equation}
\Delta\Gamma_s=\Delta\Gamma_s^{\rm SM}\cos\phi_s,
\end{equation}
where $\Delta\Gamma_s^{\rm SM}$ is negative \cite{BeLe}. As is well known, 
$\phi_s$ can be determined through $B_s\to J/\psi \phi$, which is very 
accessible at hadronic $B$-decay experiments \cite{LHC-Report,TEV-Report}. 
Strategies to determine $\phi_s$ {\it unambiguously} were proposed in 
\cite{DFN,RF-BDf-1}.

In the case of the $B_s\to D\eta^{(')}, D\phi, ...$\ modes -- the 
colour-suppressed counterparts of the $B_s\to D_s\overline{u}_s$ 
channels, untagged rates for processes where the neutral $D$ 
mesons are observed through their decays into CP eigenstates $f_\pm$ 
provide a very useful ``untagged'' rate asymmetry $\Gamma_{\pm}$, allowing
efficient and essentially unambiguous determinations of $\gamma$ from
mixing-induced observables \cite{RF-BDf-1,RF-BDf-2}. These strategies, which 
can also be implemented for $B_d\to D K_{\rm S(L)}$ modes, have 
certain similarities with those provided by (\ref{extr-1}) and 
(\ref{extr-2}). However, they do not rely on a sizeable value of 
$\Delta\Gamma_q$, as $\Gamma_{\pm}$ is extracted from ``unevolved'' 
untagged rates, which are also very useful for the analysis of 
$B_q\to D_q\overline{u}_q$ modes, as we shall see below. Since these decays 
involve charged $D_q$ mesons, the $\Gamma_{\pm}$ observable has 
unfortunately no counterpart for the colour-allowed transitions.

\boldmath
\subsection{Employing Untagged Rates in the Case of Negligible 
$\Delta\Gamma_q$}\label{subsec-DGam-0}
\unboldmath
Even for a vanishingly small width difference $\Delta\Gamma_q$, the untagged
rate (\ref{untagged}) provides valuable information, as it still allows us 
to determine the ``unevolved'', untagged rate
\begin{equation}\label{unevol-untagged}
\langle\Gamma(B_q\to D_q\overline{u}_q)\rangle\equiv
\Gamma(B^0_q\to D_q\overline{u}_q)+
\Gamma(\overline{B^0_q}\to D_q\overline{u}_q).
\end{equation}
Using (\ref{A-BbarDubar}) and (\ref{A-BDubar}), as well as
(\ref{ampl-BbarDbaru}) and (\ref{ampl-BDu}), we obtain
\begin{equation}
\frac{\langle\Gamma(B_q\to D_q\overline{u}_q)\rangle}{\Gamma(B^0_q\to 
D_q\overline{u}_q)}=1+\frac{1}{x_q^2}=
\frac{\langle\Gamma(B_q\to \overline{D}_q u_q)\rangle}{\Gamma(\overline{B^0_q}
\to\overline{D}_qu_q)}.
\end{equation}
If we now employ
\begin{equation}
\Gamma(B^0_q\to D_q\overline{u}_q)=
\Gamma(\overline{B^0_q}\to\overline{D}_qu_q),
\end{equation}
which follows from (\ref{A-BDubar}) and (\ref{ampl-BbarDbaru}),
we may write
\begin{equation}\label{xq-det1}
x_q=\eta_q\left[
\frac{\langle\Gamma(B_q\to D_q\overline{u}_q)\rangle+
\langle\Gamma(B_q\to \overline{D}_q u_q)
\rangle}{\Gamma(B^0_q\to D_q\overline{u}_q)+
\Gamma(\overline{B^0_q}\to\overline{D}_qu_q)}-1\right]^{-\frac{1}{2}},
\end{equation}
offering a very attractive ``untagged'' alternative to (\ref{xq-def-conv}), 
provided we fix the sum of the $B^0_q\to D_q\overline{u}_q$ rate 
and its CP conjugate in an efficient manner. To this end, we may replace 
the spectator quark $q$ by an up quark, which will allow us to determine 
this quantity from the CP-averaged rate of a charged $B$-meson decay as 
follows:\footnote{For simplicity, we neglect tiny
phase-space effects, which can be straightforwardly included.}
\begin{equation}\label{det-1}
\Gamma(B^0_q\to D_q\overline{u}_q)+
\Gamma(\overline{B^0_q}\to \overline{D}_q u_q)=2\,{\cal C}_q^2\left[
\Gamma(B^+\to D_q\overline{u}_u)+\Gamma(B^-\to\overline{D}_q u_u)
\right],
\end{equation}
where $u_u\in\{\pi^0,\rho^0, ...\}$ depends on the choice of $u_q$. 
For example, we have $u_u=\pi^0$ for $u_d=\pi^+$ or $u_s=K^+$, whereas
$u_u=\rho^0$ for $u_d=\rho^+$ or $u_s=K^{\ast+}$. The factor of 2 
takes into account the $1/\sqrt{2}$ factor of the $u_u$ wave function, 
and the deviation of ${\cal C}_q$ from 1 is governed by 
flavour-symmetry-breaking effects, which originate from the 
replacement of the spectator quark $q$ through an up quark. 

Since $B^+\to D_d\overline{u}_u$ is related to $B^0_d\to D_d\overline{u}_d$
through $SU(2)$ isospin arguments, we obtain to a good approximation
\begin{equation}\label{Cd-iso}
{\cal C}_d=1.
\end{equation}
In addition to the ``conventional'' isospin-breaking effects, exchange 
topologies, which contribute to $B^0_d\to D_d\overline{u}_d$ but have 
no counterpart in $B^+\to D_d\overline{u}_u$, and annihilation topologies, 
which arise only in $B^+\to D_d\overline{u}_u$ but not in 
$B^0_d\to D_d\overline{u}_d$, are another limiting factor of the 
theoretical accuracy of (\ref{Cd-iso}). Although these contributions are 
na\"\i vely expected to be very small, they may -- in principle -- be 
enhanced through rescattering processes. Fortunately, we may probe their
importance experimentally. In the case of $B_d\to D^{(\ast)\pm}\pi^\mp$
and $B^+\to D^{(\ast)+}\pi^0$ this can be done with the help of
$B_d\to D_s^{(\ast)\pm}K^\mp$ and $B^+\to D^{(\ast)+}K^0$ processes, 
respectively.

Applying (\ref{det-1}) to the $q=s$ case, we have to employ the $SU(3)$ 
flavour symmetry. If we neglect non-factorizable $SU(3)$-breaking effects, 
the ${\cal C}_s$ are simply given by appropriate form-factor ratios; 
important examples are the following ones:
\begin{equation}\label{Cs-12}
B_s\to D_s^\pm K^\mp:  \frac{F^{(0)}_{B_sK^\pm}(M_{D_s}^2)
(M_{B_s}^2-M_{K^\pm}^2)}{F^{(0)}_{B^\pm\pi^0}(M_{D_s}^2)
(M_{B_u}^2-M_{\pi^\pm}^2)}, \quad
B_s\to D_s^{\ast\pm} K^\mp: 
\frac{F^{(1)}_{B_sK^\pm}(M_{D_s^\ast}^2)}{F^{(1)}_{B^\pm\pi^0}
(M_{D_s^\ast}^2)}.
\end{equation}
Also here, we have to deal with exchange topologies, which contribute to 
$B^0_s\to D_s^{(\ast)+}K^-$ but have no counterpart in 
$B^+\to D_s^{(\ast)+}\pi^0$. Experimental probes for these topologies are 
provided by $B_s\to D^{(\ast)\pm}\pi^\mp$ processes.

As an alternative to (\ref{det-1}), we may use
\begin{equation}\label{SU3-3}
\Gamma(B^0_d\to D^{(\ast)+}\pi^-)+\Gamma(\overline{B^0_d}\to 
D^{(\ast)-}\pi^+)=\zeta\left[\Gamma(B^0_d\to D_s^{(\ast)+}\pi^-)+
\Gamma(\overline{B^0_d}\to D_s^{(\ast)-}\pi^+)\right]
\end{equation}
and
\begin{equation}\label{SU3-4}
\Gamma(B^0_s\to D_s^{(\ast)+}K^-)+
\Gamma(\overline{B^0_s}\to D_s^{(\ast)-}K^+)=\frac{1}{\zeta}
\left[\Gamma(B^0_s\to D^{(\ast)+}K^-)+
\Gamma(\overline{B^0_s}\to D^{(\ast)-}K^+)\right],
\end{equation}
where
\begin{equation}
\zeta\equiv \left(\frac{\lambda^2}{1-\lambda^2}\right)
\left[\frac{f_{D_d^{(\ast)}}}{f_{D_s^{(\ast)}}}\right]^2
\end{equation}
takes into account factorizable $SU(3)$-breaking corrections through
the ratio of the $D_d^{(\ast)}$ and $D_s^{(\ast)}$ decay constants. 
The decays on the right-hand sides of (\ref{SU3-3}) and (\ref{SU3-4}) 
have the advantage of involving ``flavour-specific'' final states $f$, 
satisfying $A(B^0_q\to f)\not=0$ and $A(\overline{B^0_q}\to f)=0$.
In this important special case, the time-dependent untagged rates take 
the following simple forms:
\begin{equation}
\langle\Gamma(B_q(t)\to f)\rangle\equiv
\Gamma(B^0_q(t)\to f)+\Gamma(\overline{B^0_q}(t)\to f)=\Gamma(B^0_q\to f)
\cosh(\Delta\Gamma_qt/2)e^{-\Gamma_qt}
\end{equation}
\begin{equation}
\langle\Gamma(B_q(t)\to\overline{f})\rangle\equiv
\Gamma(B^0_q(t)\to \overline{f})+\Gamma(\overline{B^0_q}(t)\to \overline{f})
=\Gamma(\overline{B^0_q}\to\overline{f})\cosh(\Delta\Gamma_qt/2)
e^{-\Gamma_qt},
\end{equation}
and allow an efficient extraction of the CP-averaged rate 
$\Gamma(B^0_q\to f)+\Gamma(\overline{B^0_q}\to\overline{f})$ with the help of
\begin{equation}\label{untagged-FS}
\langle\Gamma(B_q(t)\to f)\rangle+\langle\Gamma(B_q(t)\to\overline{f})\rangle=
\left[\Gamma(B^0_q\to f)+\Gamma(\overline{B^0_q}\to\overline{f})\right]
\cosh(\Delta\Gamma_qt/2)e^{-\Gamma_qt}.
\end{equation}
Obviously, in the case of $q=d$, (\ref{det-1}) is theoretically cleaner than 
(\ref{SU3-3}), providing -- in combination with (\ref{xq-det1}) -- a very
interesting avenue to determine $x_d$. On the other hand, the modes on the 
right-hand side of (\ref{SU3-3}) are more accessible from an experimental 
point of view, and were already observed at the $B$ factories 
\cite{BDspi-measurements}. 

Since simple colour-transparency arguments do not apply to 
$B^0_q\to D_q\overline{u}_q$, $B^+\to D_q\overline{u}_u$ modes, as we 
noted in Subsection~\ref{subsec:fact}, expressions (\ref{Cs-12}),
(\ref{SU3-3}) and (\ref{SU3-4}) may receive sizeable non-factorizable 
$SU(3)$-breaking corrections. However, there is yet another possibility 
to exploit (\ref{unevol-untagged}). To this end, we factor out the 
$B^0_q\to \overline{D}_qu_q$ rate, where
factorization is expected to work well \cite{fact}, yielding
\begin{equation}\label{unevol-untagged-2}
\frac{\langle\Gamma(B_q\to D_q\overline{u}_q)\rangle}{\Gamma(\overline{B^0_q}
\to D_q \overline{u}_q)}=1+x_q^2=
\frac{\langle\Gamma(B_q\to \overline{D}_q u_q)\rangle}{\Gamma(B^0_q
\to \overline{D}_q u_q)},
\end{equation}
which implies
\begin{equation}\label{xs-def-gol}
x_q=\eta_q\sqrt{\frac{\langle\Gamma(B_q\to D_q\overline{u}_q)\rangle+
\langle\Gamma(B_q\to \overline{D}_q u_q)\rangle}{\Gamma(\overline{B^0_q}\to 
D_q\overline{u}_q)+\Gamma(B^0_q\to \overline{D}_q u_q)}-1}.
\end{equation}
In the $q=d$ case, it will -- in analogy to (\ref{Cm-def}) -- be
impossible to resolve the vanishingly small $x_q^2$ term in 
(\ref{unevol-untagged-2}). On the other hand, this may well be possible
in the $q=s$ case. If we use
\begin{eqnarray}
\lefteqn{\Gamma(\overline{B^0_s}\to D_s^{(\ast)+}K^-)+
\Gamma(B^0_s\to D_s^{(\ast)-}K^+)}\nonumber\\
&&=\left(\frac{\lambda^2}{1-\lambda^2}\right)\left(\frac{f_K}{f_\pi}\right)^2
\left[\Gamma(\overline{B^0_s}\to D_s^{(\ast)+}\pi^-)+
\Gamma(B^0_s\to D_s^{(\ast)-}\pi^+)\right],\label{BsDsK-det}
\end{eqnarray}
expression (\ref{xs-def-gol}) offers a very attractive possibility to 
determine the values of $x_{s(\ast)}$, where $(f_K/f_\pi)^2$ describes 
factorizable $SU(3)$-breaking effects. Additional corrections are due to 
exchange topologies, which arise in $\overline{B^0_s}\to D_s^{(\ast)+}K^-$, 
but are not present in $\overline{B^0_s}\to D_s^{(\ast)+}\pi^-$. However, 
as we already noted, their contributions are expected to be very small, 
and can be probed experimentally through $B_s\to D^{(\ast)\pm}\pi^\mp$ 
processes. Since the $\overline{B^0_s}\to D_s^{(\ast)+}\pi^-$ and
$B^0_s\to D_s^{(\ast)-}\pi^+$ rates involve flavour-specific final states, 
we may efficiently determine their sum from untagged $B_s$ data samples, 
with the help of (\ref{untagged-FS}). In this context, it should also be
noted that these rates are enhanced by a factor of 
$(1-\lambda^2)/\lambda^2\approx20$ with respect to the
$B_s\to D_s^{(\ast)\pm}K^\mp$ rates. Moreover, non-factorizable effects 
are expected to play a minor r\^ole in (\ref{BsDsK-det}) because of 
colour-transparency arguments, in contrast to (\ref{Cs-12}) and 
(\ref{SU3-4}). Further calculations along \cite{fact} should provide an 
even more accurate treatment of the $SU(3)$-breaking corrections.
In comparison with (\ref{xq-def-conv}), the advantage of the strategy
offered by (\ref{xs-def-gol}) and (\ref{BsDsK-det}) is the use of untagged 
rates, which are particularly promising in terms of efficiency, acceptance 
and purity, and do not require the measurement of the time-dependent 
$\cos(\Delta M_st)$ terms in (\ref{rate-asym}). Interestingly, the
quantity $1+x_s^2$, which can nicely be determined through the combination 
of (\ref{xs-def-gol}) and (\ref{BsDsK-det}), will play an important 
r\^ole in Section~\ref{sec:comb}.

As we have seen above, the untagged rates introduced in (\ref{untagged}) 
provide various strategies to determine the hadronic parameters $x_q$, 
some of which are particularly favourable. In order to implement 
these approaches, we must not rely on a sizeable width difference 
$\Delta\Gamma_q$. It will be interesting to see whether they will 
eventually yield a consistent picture of the $x_q$. Following these 
lines, we may also obtain valuable insights into hadron dynamics.

\boldmath
\section{Bounds on $\phi_q+\gamma$}\label{sec:bounds}
\unboldmath
\setcounter{equation}{0}
If we keep $x_q$ and $\delta_q$ as ``unknown'', i.e.\ free parameters 
in (\ref{Sp-def}) and (\ref{Sm-def}), we may derive the following bounds:
\begin{equation}\label{bound-s-1}
|\sin(\phi_q+\gamma)|\geq|\langle S_q\rangle_+|
\end{equation}
\begin{equation}\label{bound-c-1}
|\cos(\phi_q+\gamma)|\geq|\langle S_q\rangle_-|.
\end{equation}
On the other hand, if we assume that $x_q$ has been determined with the 
help of the ``untagged'' strategies proposed in
Subsection~\ref{subsec-DGam-0}, we may fix the quantities $s_+$ and $s_-$ 
introduced in (\ref{s-p}) and (\ref{s-m}), respectively, providing more 
stringent constraints:
\begin{equation}\label{bound-s-2}
|\sin(\phi_q+\gamma)|\geq|s_+|
\end{equation}
\begin{equation}\label{bound-c-2}
|\cos(\phi_q+\gamma)|\geq|s_-|.
\end{equation}
Interestingly, (\ref{bound-s-1}) and (\ref{bound-s-2}) allow us to
exclude a certain range of values of $\phi_q+\gamma$ around $0^\circ$
and $180^\circ$, whereas (\ref{bound-c-1}) and (\ref{bound-c-2}) 
provide complementary information, excluding a certain range around 
$90^\circ$ and $270^\circ$. The constraints in (\ref{bound-s-1}) and 
(\ref{bound-c-1}) have the advantage of not requiring knowledge of $x_q$. 
On the other hand, because of the small value of $x_d$, we may only 
expect useful information from them in the case of $q=s$. Once $s_+$ and 
$s_-$ have been extracted, it is of course also possible to determine 
$\sin^2(\phi_q+\gamma)$ through the complicated expression in 
(\ref{sin2phi-det}), as discussed in Section~\ref{sec:conv}. However, 
since the resulting values for $\phi_q+\gamma$ suffer from multiple 
discrete ambiguities, the information they are expected to provide 
about this phase is -- in general -- not significantly better than the 
constraints following from the very simple relations in (\ref{bound-s-2}) 
and (\ref{bound-c-2}).

\begin{table}
\begin{center}
\begin{tabular}{|c||c|c|c|c|c|c|}
\hline
$q$ & $S_q$ & $\overline{S}_q$  & $\langle S_q\rangle_+$ & 
$\langle S_q\rangle_-$ & ~~~$s_+$~~~ & ~~~$s_-$~~~ \\ 
\hline
\hline
$d$ & $-3.89\%$ & $-3.89\%$ & $-3.89\%$ & $+0.00\%$ & $+95.6\%$ & $+0.00\%$ \\
$s$ & $+59.7\%$ & $+59.7\%$ & $+59.7\%$ & $+0.00\%$ & $+86.6\%$ & $+0.00\%$ \\
\hline
$d$ & $-2.22\%$ & $-3.74\%$ & $-2.98\%$ & $-0.76\%$ & $+73.3\%$ & $+18.8\%$ \\
$s$ & $+67.9\%$ & $+23.6\%$ & $+45.8\%$ & $-22.2\%$ & $+66.3\%$ & $-32.1\%$ \\
\hline
\end{tabular}
\caption{The mixing-induced observables in the case of $L=0$, 
$\gamma=60^\circ$, $\phi_d=47^\circ$, $\phi_s=0^\circ$, $R_b=0.4$ and 
$a_q=1$: the upper half corresponds to factorization, i.e.\ 
$\delta_q=0^\circ$, whereas the lower half illustrates a 
non-factorization scenario with $\delta_q=40^\circ$. Note that we have 
$\langle C_s\rangle_-=0.724$, while the deviation of 
$\langle C_d\rangle_-$ from 1 is negligibly small.}\label{tab:illu}
\end{center}
\end{table}

It is instructive to illustrate this feature with the help of a few numerical 
examples. To this end, we assume $\gamma=60^\circ$, $\phi_d=47^\circ$ and 
$\phi_s=0^\circ$, which would be in perfect agreement with the Standard 
Model, as well as $R_b=0.4$ and $a_q=1$. Let us consider the decays 
$B_d\to D^\pm\pi^\mp$ and $B_s\to D_s^\pm K^\mp$, which have $L=0$. As far 
as $\delta_q$ is concerned, we may then distinguish between a 
``factorization'' scenario with $\delta_q=0^\circ$ (see (\ref{phase-fact})), 
and a ``non-factorization'' scenario, corresponding to $\delta_q=40^\circ$. 
For simplicity, we shall use the same hadronic parameters
$a_qe^{i\delta_q}$ for the $q=d$ and $q=s$ cases. The corresponding  
mixing-induced observables are listed in Table~\ref{tab:illu}. Let us also 
assume that $\phi_d$ and $\phi_s$ will be unambiguously known by the time 
these observables can be measured. As we have already noted, because of
the small value of $x_d$, (\ref{bound-s-1}) and (\ref{bound-c-1}) do not
provide non-trivial constraints on $\phi_d+\gamma$, in contrast to their
application to the $q=s$ case. 

Let us first focus on the factorization scenario, corresponding to the
upper half of Table~\ref{tab:illu}. Since $\langle S_q\rangle_-$ 
and $s_-$ vanish in this case, as these observable combinations are 
proportional to $\sin\delta_q$, (\ref{bound-c-1}) and (\ref{bound-c-2}) 
imply only trivial constraints on $\phi_q+\gamma$. However, we may 
nevertheless obtain interesting bounds in this case. For the $q=d$ example,
the situation is as follows: if we employ (\ref{sgn-cos-del}) and take 
into account that $x_d$ is negative, the negative sign of 
$\langle S_d\rangle_+$ implies a positive value of 
$\sin(\phi_d+\gamma)$, i.e.\ $0^\circ\leq \phi_d+\gamma \leq 180^\circ$. 
Applying now (\ref{bound-s-2}), we obtain 
$73^\circ\leq\phi_d+\gamma\leq107^\circ$ from $s_+$, which corresponds to 
$26^\circ\leq\gamma\leq60^\circ$, providing valuable information about 
$\gamma$. On the other hand, if we use again that $\sin(\phi_d+\gamma)$
is positive, the complicated expression (\ref{sin2phi-det}) implies the 
threefold solution $\gamma=26^\circ\lor43^\circ\lor60^\circ$, which covers 
essentially the whole range following from the simple relation in 
(\ref{bound-s-2}). It is very interesting to complement the information on 
$\gamma$ thus obtained from $B_d\to D^\pm\pi^\mp$ with the one provided by 
its $B_s\to D_s^\pm K^\mp$ counterpart. Using again (\ref{sgn-cos-del}), 
the positive sign of $\langle S_s\rangle_+$ implies that 
$\sin(\phi_s+\gamma)$ is positive, i.e.\ 
$0^\circ\leq \phi_s+\gamma \leq 180^\circ$. We may now apply (\ref{bound-s-1}) 
to obtain the bound $37^\circ\leq \phi_s+\gamma\leq 143^\circ$ from 
$\langle S_s\rangle_+$; a narrower range follows from $s_+$ through 
(\ref{bound-s-2}), and is given by 
$60^\circ\leq \phi_s+\gamma \leq 120^\circ$. Since $\phi_s=0^\circ$, 
we may identify these ranges directly with bounds on $\gamma$. On
the other hand, the complicated expression (\ref{sin2phi-det}) implies
the threefold solution $\gamma=60^\circ\lor90^\circ\lor120^\circ$, which 
falls perfectly into the range provided by $s_+$, which can be obtained 
in a much simpler manner. We now make the very interesting observation that 
the $q=s$ range of $60^\circ\leq\gamma \leq 120^\circ$ is highly 
complementary to its $q=d$ counterpart of $26^\circ\leq\gamma\leq60^\circ$, 
leaving $60^\circ$ as the only overlap. Consequently, in this example, 
the combination of our simple bounds on $\phi_d+\gamma$ and $\phi_s+\gamma$ 
yields the single solution of $\gamma=60^\circ$, which corresponds to our
input value, thereby nicely demonstrating the potential power of these 
constraints. 

Let us now perform the same exercise for the non-factorization scenario,
represented by the lower half of Table~\ref{tab:illu}. In the case of 
$q=d$, $s_+$ and $s_-$ imply $47^\circ\leq\phi_d+\gamma\leq133^\circ$ 
and $(0^\circ\leq\phi_d+\gamma\leq79^\circ) \lor (101^\circ\leq
\phi_d+\gamma\leq180^\circ)$, respectively, which can be combined with
each other, taking also $\phi_d=47^\circ$ into account, to obtain 
$(0^\circ\leq\gamma\leq32^\circ) \lor (54^\circ\leq \gamma\leq86^\circ)$. 
On the other hand, if we apply (\ref{sin2phi-det}) and use that 
$\sin(\phi_d+\gamma)$ is positive, we obtain the fourfold solution
$\gamma=3^\circ \lor 26^\circ \lor 60^\circ \lor 83^\circ$. Let us now
consider the $q=s$ case. Here $\langle S_s\rangle_+$ and 
$\langle S_s\rangle_-$ imply $27^\circ\leq\phi_s+\gamma\leq153^\circ$ and
$(0^\circ\leq\phi_s+\gamma\leq77^\circ) \lor (103^\circ\leq\phi_s+\gamma\leq
180^\circ)$, respectively, yielding the combined range 
$(27^\circ\leq\phi_s+\gamma\leq77^\circ) \lor (103^\circ\leq\phi_s+\gamma\leq
153^\circ)$. Using $s_+$ and $s_-$, and taking into account that 
$\phi_s=0^\circ$, we obtain the more stringent constraint 
$(42^\circ\leq\gamma\leq71^\circ) \lor (109^\circ\leq\gamma\leq138^\circ)$,
whereas (\ref{sin2phi-det}) would imply the fourfold solution
$\gamma=50^\circ \lor 60^\circ \lor 120^\circ \lor 130^\circ$, providing
essentially the same information. We observe again that the bounds on 
$\gamma$ arising in the $q=d$ and $q=s$ cases are highly complementary 
to each other, having a small overlap of $54^\circ\leq\gamma\leq71^\circ$. 
Although the constraint on $\gamma$ following from the bounds on 
$\phi_q+\gamma$ would now not be as sharp as in the factorization scenario 
discussed above, this approach would still provide very non-trivial 
information about this particularly important angle of the unitarity 
triangle.

In Table~\ref{tab:illu}, we have considered a Standard-Model-like
scenario for the weak phases. However, as argued in \cite{FIM}, the present 
data are also perfectly consistent with the picture of 
$(\phi_d,\gamma)=(133^\circ,120^\circ)$, corresponding to 
new-physics contributions to $B^0_d$--$\overline{B^0_d}$ mixing.
Since we have $\sin(\phi_d+\gamma)\to -\sin(\phi_d+\gamma)$
for $\phi_d\to 180^\circ-\phi_d$, $\gamma\to 180^\circ-\gamma$,
the sign of the $(-1)^L\langle S_d\rangle_+$ observable combination
allows us to distinguish between the $(\phi_d,\gamma)=(47^\circ,60^\circ)$
and $(133^\circ,120^\circ)$ scenarios, corresponding to 
$\sin(\phi_d+\gamma)=+0.956$ and $-0.956$, respectively. Practically,
this can be done with the help of $B_d\to D^{(\ast)\pm}\pi^\mp$ modes.
If we take into account that the $x_{d(\ast)}$ are negative, include 
properly the $(-1)^L$ factors and fix the signs of $\cos\delta_{d(\ast)}$ 
through (\ref{sgn-cos-del}), we find that a positive value of the 
$\langle S_{d(\ast)}\rangle_+$ observables would be in favour of the 
``unconventional'' $(\phi_d,\gamma)=(133^\circ,120^\circ)$ scenario,
whereas a negative value would point towards the Standard-Model picture
of $(\phi_d,\gamma)=(47^\circ,60^\circ)$.
A first preliminary analysis of $B_d\to D^{\ast\pm}\pi^\mp$ by the BaBar 
collaboration \cite{BaBar-BDpi} gives 
\begin{equation}
\langle S_{d\ast}\rangle_+=-0.063\pm0.024 \, (\mbox{stat.})
\pm0.017 \, (\mbox{syst.}),
\end{equation}
thereby favouring the latter case.

\boldmath
\section{Combined Analysis of $B_{s,d}\to D_{s,d}\overline{u}_{s,d}$ 
Modes}\label{sec:comb}
\unboldmath
\setcounter{equation}{0}
As we have seen in the previous section, it is very useful to make a
simultaneous analysis of $B_s\to D_s\overline{u}_s$ and 
$B_d\to D_d\overline{u}_d$ decays. Let us now further explore this 
observation. Using (\ref{Sp-def}) and (\ref{Sm-def}), we may write
\begin{equation}\label{cos-rel}
\left[\frac{a_s\cos\delta_s}{a_d\cos\delta_d}\right]R=
-(-1)^{L_s-L_d}\left[\frac{\sin(\phi_d+\gamma)}{\sin(\phi_s+\gamma)}\right]
\left[\frac{\langle S_s\rangle_+}{\langle S_d\rangle_+}\right]
\end{equation}
and
\begin{equation}\label{sin-rel}
\left[\frac{a_s\sin\delta_s}{a_d\sin\delta_d}\right]R=
-(-1)^{L_s-L_d}\left[\frac{\cos(\phi_d+\gamma)}{\cos(\phi_s+\gamma)}\right]
\left[\frac{\langle S_s\rangle_-}{\langle S_d\rangle_-}\right],
\end{equation}
respectively, where
\begin{equation}
R\equiv \left(\frac{1-\lambda^2}{\lambda^2}\right)
\left[\frac{1+x_d^2}{1+x_s^2}\right].
\end{equation}
Using the results derived in Subsection~\ref{subsec-DGam-0}, we may
easily determine the parameter $R$, where the $x_d^2$ 
term is negligibly small, and $x_s$ enters only through $1+x_s^2$, i.e.\
a moderate correction. To be specific, let us consider the 
$B_s\to D_s^{(\ast)\pm}K^\mp$ channels. If we insert (\ref{BsDsK-det}) 
into (\ref{xs-def-gol}), we arrive at
\begin{equation}\label{R-det}
R_{(\ast)}=\left(\frac{f_K}{f_\pi}\right)^2 
\left[\frac{\Gamma(\overline{B^0_s}\to D_s^{(\ast)+}\pi^-)+
\Gamma(B^0_s\to D_s^{(\ast)-}\pi^+)}{\langle\Gamma(B_s\to D_s^{(\ast)+}K^-)
\rangle+\langle\Gamma(B_s\to D_s^{(\ast)-}K^+)\rangle}\right],
\end{equation}
where the decay rates can be straightforwardly extracted from untagged 
$B_s$ data samples with the help of (\ref{untagged}) and (\ref{untagged-FS}). 
As we have emphasized in Subsection~\ref{subsec-DGam-0}, non-factorizable 
$SU(3)$-breaking corrections to this relation are expected to be very small. 

If we look at Fig.~\ref{fig:BqDu}, we see that each $B_s\to D_s\overline{u}_s$ 
mode has a counterpart $B_d\to D_d\overline{u}_d$, which can be obtained from 
the $B_s$ transition by simply replacing all strange quarks through down 
quarks; an important example is the $B_s^0\to D_s^{(\ast)+}K^-$, 
$B_d^0\to D^{(\ast)+}\pi^-$ system. For such decay pairs, we 
have $L_s=L_d$, and the $U$-spin flavour symmetry of strong interactions, 
which relates strange and down quarks in the same manner as ordinary isospin 
relates up and down quarks, implies the following relations for the
corresponding hadronic parameters:\footnote{Note that these relations do 
not rely on the neglect of (tiny) exchange topologies.}
\begin{equation}\label{U-spin-rel}
a_s=a_d, \quad \delta_s=\delta_d,
\end{equation}
which we may apply in a variety of ways. 

Let us first consider a factorization-like scenario, where 
$\cos\delta_s\approx\pm1\approx\cos\delta_d$ and $\langle S_s\rangle_-\approx
0\approx\langle S_d\rangle_-$ (see Table~\ref{tab:illu}). In this case, 
(\ref{sin-rel}) would not be applicable. However, we may use 
(\ref{cos-rel}) to determine $\tan\gamma$ through
\begin{equation}\label{tan-gam-det-C}
\tan\gamma=-\left[\frac{\sin\phi_d-S\sin\phi_s}{\cos\phi_d-S\cos\phi_s}
\right]\stackrel{\phi_s=0^\circ}{=}
-\left[\frac{\sin\phi_d}{\cos\phi_d-S}\right],
\end{equation}
where
\begin{equation}
\left.S\right|_{U{\rm \,spin}}=-R\left[\frac{\langle S_d\rangle_+}{\langle 
S_s\rangle_+}\right].
\end{equation}
If we follow these lines, we obtain a twofold solution 
$\gamma=\gamma_1\lor\gamma_2$, where we may choose 
$\gamma_1\in[0^\circ,180^\circ]$ and $\gamma_2=\gamma_1+180^\circ$; the
theoretical uncertainty would mainly be limited by $U$-spin-breaking
corrections to $a_s=a_d$, apart from tiny corrections to 
$\cos\delta_s=\cos\delta_d$. If we assume -- as is usually done -- that 
$\gamma$ lies between $0^\circ$ 
and $180^\circ$, as is implied by the Standard-Model interpretation of
$\varepsilon_K$, which measures the ``indirect'' CP violation in the 
neutral kaon system, we may immediately exclude the $\gamma_2$ solution. 
However, since $\varepsilon_K$ may well be affected by new physics, it
is desirable to check whether $\gamma$ actually falls in the interval 
$[0^\circ,180^\circ]$. To this end, we may use (\ref{sgn-cos-del}) 
and the signs of the $\langle S_q\rangle_+$ observables, as we
have seen in the examples discussed in Section~\ref{sec:bounds}. 

\begin{figure}
\centerline{\rotate[r]{
\epsfysize=10.9truecm
{\epsffile{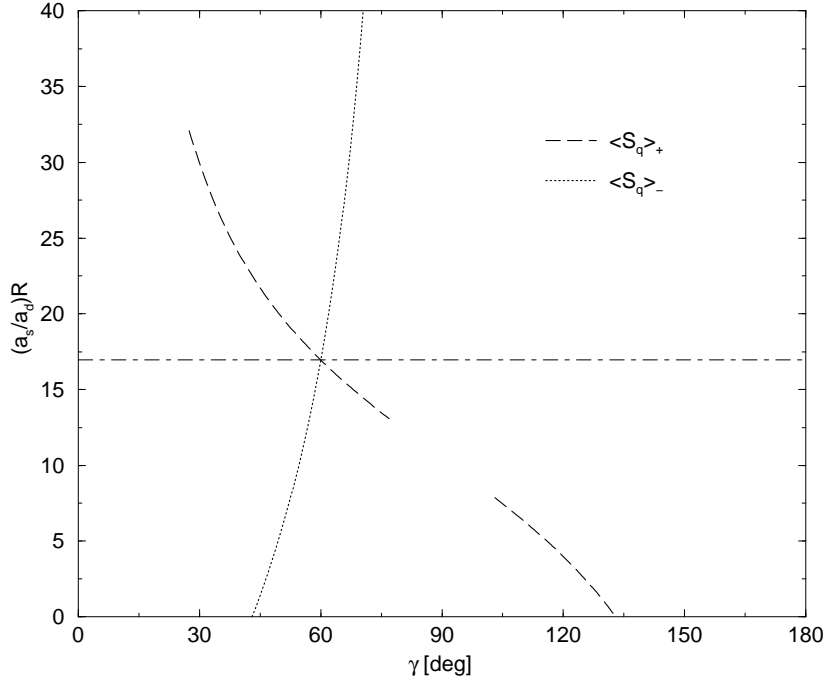}}}}
\caption{Extraction of $\gamma$ assuming $\delta_s=\delta_d$ for the 
non-factorization scenario in Table~\ref{tab:illu}; the dashed and 
dotted curves were calculated with the help of (\ref{cos-rel}) and 
(\ref{sin-rel}), respectively.}\label{fig:extr-gam1}
\end{figure}

Let us now consider a non-factorization-like scenario with sizeable 
CP-conserving strong phases, so that we may also employ (\ref{sin-rel}), 
as the $\langle S_q\rangle_-$ observables would no longer vanish. 
If we assume that $\delta_s=\delta_d$, we may calculate $(a_s/a_d)R$ both 
with the help of the $\langle S_q\rangle_+$ observables through 
(\ref{cos-rel}) and with the help of the $\langle S_q\rangle_-$ observables 
through (\ref{sin-rel}). The intersection of the corresponding curves then 
fixes $\gamma$ and $(a_s/a_d)R$. Comparing the value of $(a_s/a_d)R$ thus 
extracted with (\ref{R-det}), we could determine $a_s/a_d$. If we use
the observables given in the lower half of Table~\ref{tab:illu}, which were 
calculated for $\delta_s=\delta_d =40^\circ$ and $a_s=a_d=1$, we obtain 
the contours shown in Fig.~\ref{fig:extr-gam1}, where we have also taken 
the bounds implied by (\ref{bound-s-1}) and (\ref{bound-c-1}) into account,
and have represented the curves originating from (\ref{cos-rel}) and 
(\ref{sin-rel}) through the dashed and dotted lines, respectively.
We observe that the intersection of these contours gives actually our
input value of $\gamma=60^\circ$, without any discrete ambiguity.
These observations can easily be put on a more formal level, since 
(\ref{cos-rel}) and (\ref{sin-rel}) imply the following {\it exact} 
relation:
\begin{equation}
\frac{\tan(\phi_d+\gamma)}{\tan(\phi_s+\gamma)}=\left[
\frac{\tan\delta_d}{\tan\delta_s}\right]
\left[\frac{\langle S_s\rangle_-}{\langle S_s\rangle_+}\right]
\left[\frac{\langle S_d\rangle_+}{\langle S_d\rangle_-}\right]
\, \stackrel{U{\rm \,spin}}{\longrightarrow} \, 
\left[\frac{\langle S_s\rangle_-}{\langle S_s\rangle_+}\right]
\left[\frac{\langle S_d\rangle_+}{\langle S_d\rangle_-}\right].
\end{equation}
Consequently, the theoretical uncertainty of the resulting value of $\gamma$
would {\it only} be limited by $U$-spin-breaking corrections to 
$\tan\delta_s=\tan\delta_d$; in Fig.~\ref{fig:extr-gam1}, they would
enter through a systematic relative shift of the dashed and dotted
contours.

\begin{figure}
\centerline{\rotate[r]{
\epsfysize=10.9truecm
{\epsffile{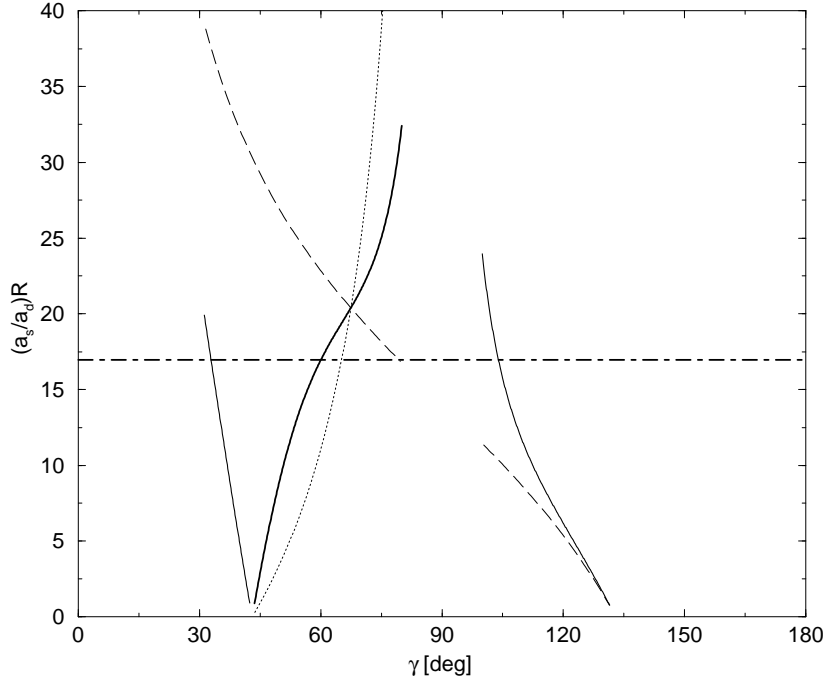}}}}
\caption{Extraction of $\gamma$ with the help of (\ref{gold-rel}), yielding
the solid lines, for an example with $\delta_d=50^\circ$ and 
$\delta_s=30^\circ$, as discussed in the text.}\label{fig:extr-gam2}
\end{figure}

Finally, we may also extract $\gamma$ {\it without} assuming that 
$\delta_s$ is equal to $\delta_d$. To this end, we use the {\it exact} 
relation
\begin{equation}\label{gold-rel}
\left(\frac{a_{s}}{a_{d}}\right)R=
\sigma\left|\frac{\sin(2\phi_d+2\gamma)}{\sin(2\phi_s+2\gamma)}\right|
\sqrt{\frac{\langle S_{s}\rangle_+^2\cos^2(\phi_s+\gamma)+
\langle S_{s}\rangle_-^2\sin^2(\phi_s+\gamma)}{\langle 
S_{d}\rangle_+^2\cos^2(\phi_d+\gamma)+
\langle S_{d}\rangle_-^2\sin^2(\phi_d+\gamma)}},
\end{equation}
where we have 
\begin{equation}\label{sigma1}
\sigma=-{\rm sgn}\left\{\langle S_{s}\rangle_+\langle S_{d}\rangle_+
\sin(\phi_d+\gamma)\sin(\phi_s+\gamma)\right\}
\end{equation}
if we assume that $\cos\delta_s$ and $\cos\delta_d$ have the same
sign, and
\begin{equation}\label{sigma2}
\sigma=-{\rm sgn}\left\{\langle S_{s}\rangle_-\langle S_{d}\rangle_-
\cos(\phi_d+\gamma)\cos(\phi_s+\gamma)\right\}
\end{equation}
if we assume that $\sin\delta_s$ and $\sin\delta_d$ have the same
sign. Using (\ref{gold-rel}), we may calculate $(a_{s}/a_{d})R$ in an 
{\it exact} manner as a function of $\gamma$ from the measured values 
of the mixing-induced observables $\langle S_{s}\rangle_\pm$ and 
$\langle S_{d}\rangle_\pm$. On the other hand, we have $a_s\approx a_d$ 
because of the $U$-spin flavour symmetry, and may efficiently fix $R$ from 
untagged $B_s$ data samples through (\ref{R-det}), allowing us to 
determine $\gamma$. Let us illustrate how this strategy works in practice 
by considering again an example, corresponding to $a_s=a_d=1$, 
$\delta_d=50^\circ$ and $\delta_s=30^\circ$. Moreover, as in 
Table~\ref{tab:illu}, we choose $\gamma=60^\circ$, $\phi_d=47^\circ$, 
$\phi_s=0^\circ$ and $R_b=0.4$, implying $\langle S_{d}\rangle_+=-2.50\%$, 
$\langle S_{d}\rangle_-=-0.91\%$, $\langle S_{s}\rangle_+=51.7\%$, 
$\langle S_{s}\rangle_-=-17.2\%$. If we apply (\ref{bound-s-1}) and 
(\ref{bound-c-1}) to the $B_s$ observables, we obtain 
$31^\circ\leq\gamma\leq80^\circ \lor 100^\circ\leq\gamma\leq133^\circ$. 
Constraining $\gamma$ to this range, the right-hand side of (\ref{gold-rel}) 
yields the solid lines shown in Fig.~\ref{fig:extr-gam2}, where we 
have represented the ``measured'' value of $R$ through the horizontal 
dot-dashed line; the three lines emerge if we fix $\sigma$ 
through (\ref{sigma1}), yielding the threefold solution 
$\gamma=33^\circ\lor60^\circ\lor 104^\circ$. However, (\ref{sigma2})
leaves only the thicker solid line in the middle, thereby implying the 
single solution $\gamma=60^\circ$. In this particular example, 
the extracted value for $\gamma$ would be quite stable with respect to
variations of $(a_s/a_d)R$, i.e.\ would not be very sensitive to 
$U$-spin-breaking corrections to $a_s=a_d$. We have also included the 
contours corresponding to (\ref{cos-rel}) and (\ref{sin-rel}) through 
the dashed and dotted curves, as in Fig.~\ref{fig:extr-gam1}; their 
intersection would now give $\gamma=68^\circ$, deviating by only 
$8^\circ$ from the ``correct'' value. It should be noted that we may 
also determine the strong phases $\delta_s$ and $\delta_d$ with the help of
\begin{equation}\label{strong-phase-det}
\tan\delta_{q}=-\left[\frac{\langle S_{q}\rangle_-}{\langle 
S_{q}\rangle_+}\right]\tan(\phi_q+\gamma),
\end{equation}
providing valuable insights into non-factorizable $U$-spin-breaking
effects. 

In comparison with the conventional $B_q\to D_q\overline{u}_q$ approaches --
apart from issues related to multiple discrete ambiguities -- the most
important advantage of the strategies proposed above is that they do 
{\it not} require the resolution of $x_q^2$ terms, since the mixing-induced 
observables $\langle S_d\rangle_\pm$ and $\langle S_s\rangle_\pm$ are 
proportional to $x_d$ and $x_s$, respectively. In particular, $x_d$ has 
{\it not} to be fixed, and $x_s$ may only enter through $1+x_s^2$, i.e.\ a 
moderate correction, which can straightforwardly be included through 
untagged $B_s$ rate analyses. Interestingly, the motivation to measure 
$x_s$ and $x_d$ accurately is here related only to the feature that these 
parameters would allow us to take into account possible $U$-spin-breaking 
corrections to (\ref{gold-rel}) through
\begin{equation}
\frac{a_s}{a_d}=\left(\frac{\lambda^2}{1-\lambda^2}\right)\left|
\frac{x_s}{x_d}\right|.
\end{equation}
After all these steps of progressive refinement, we would eventually 
obtain a theoretically clean value of $\gamma$.

For a theoretical discussion of the $U$-spin-breaking effects affecting the
ratio $a_s/a_d$, we may distinguish -- apart from mass factors -- between 
two pieces, 
\begin{equation}\label{as-ad-ratio}
\frac{a_{s}}{a_{d}}\sim \zeta_1
\times \zeta_2,
\end{equation}
which can be written for the $B_s\to D_s^{(\ast)\pm}
K^\mp$, $B_d\to D^{(\ast)\pm}\pi^\mp$ system -- if we apply the factorization 
approximation --  with the help of (\ref{as-fact-HQ}), (\ref{ad-fact-HQ}) and 
(\ref{as-ast-fact-HQ}), (\ref{ad-ast-fact-HQ}) as follows:
\begin{equation}\label{zeta1-fact}
\left.\zeta_1^{(\ast)}\right|_{\rm fact}=
\frac{f_{\pi^\pm}\xi_d(w_d^{(\ast)})}{f_{K^\pm}
\xi_s(w_s^{(\ast)})}
\end{equation}
\begin{equation}\label{zeta2-fact}
\left.\zeta_2\right|_{\rm fact}
=\frac{f_{D_s}F^{(0)}_{B_sK^\pm}(M_{D_s}^2)}{f_{D_d}
F^{(0)}_{B_d\pi^\pm}(M_{D_d}^2)}, \quad
\left.\zeta_2^{\ast}\right|_{\rm fact}
=\frac{f_{D_s^\ast}F^{(1)}_{B_sK^\pm}
(M_{D_s^\ast}^2)}{f_{D_d^\ast}F^{(1)}_{B_d\pi^\pm}(M_{D_d^\ast}^2)}.
\end{equation}
Because of the arguments given in Subsection~\ref{subsec:fact}, the 
factorized expression (\ref{zeta1-fact}) for $\zeta_1^{(\ast)}$ is expected 
to work well. Studies of the light-quark dependence of the Isgur--Wise 
function were performed in \cite{xi-SU3} within heavy-meson chiral 
perturbation theory, indicating an enhancement of $\xi_s/\xi_d$ at the 
level of $5\%$. The application of the same formalism to $f_{D_s}/f_{D_d}$ 
yields values at the 1.2 level \cite{fD-SU3}, which is of the same order of 
magnitude as recent lattice calculations (see, for example, \cite{lattice}). 
In the case of $\zeta_2^{(\ast)}$, (\ref{zeta2-fact}) may receive sizeable 
non-factorizable corrections, since simple colour-transparency arguments are 
not on solid ground, and the new theoretical developments related to 
factorization that were presented in \cite{fact} are not 
applicable. Moreover, we are not aware of quantitative studies of the 
$SU(3)$-breaking effects arising from the $B_s\to K^\pm$, $B_d\to \pi^\pm$ 
form factors in (\ref{zeta2-fact}), which could be done, for instance, 
with the help of lattice or sum-rule techniques. Following the latter 
approach, sizeable $SU(3)$-breaking corrections were found for the 
$B_s\to K^{\ast\pm}$, $B_d\to \rho^{\pm}$ form factors in \cite{BB}. 
Hopefully, a better theoretical treatment of the $U$-spin-breaking 
corrections to $a_s/a_d$ will be available by the time the 
$B_q\to D_q\overline{u}_q$ measurements can be performed in practice. 

The new strategies proposed above complement other $U$-spin approaches 
to extract $\gamma$ \cite{U-Spin,RF-BsKK}, where the $U$-spin-related 
$B_s\to K^+K^-$, $B_d\to\pi^+\pi^-$ system is particularly promising
\cite{LHC-Report,TEV-Report,RF-BsKK}. Since penguin topologies play here 
a crucial r\^ole, whereas these topologies do not contribute to the 
$B_s\to D_s^{(\ast)\pm}K^\mp$, $B_d\to D^{(\ast)\pm}\pi^\mp$ system, 
it will be very interesting to see whether inconsistencies for $\gamma$ 
will emerge from the data.

\section{Conclusions}\label{sec:concl}
\setcounter{equation}{0}
Let us now summarize the main points of our analysis:
\begin{itemize}
\item We have shown that $B_s\to D_s^{\pm}K^\mp, D_s^{\ast\pm}K^\mp, ...$\ 
and $B_d\to D^{\pm}\pi^\mp, D^{\ast\pm}\pi^\mp, ...$\ decays can be 
described through the same set of formulae by just making straightforward 
replacements of variables. We have also pointed out that a factor of 
$(-1)^L$ arises in the expressions for the mixing-induced observables. 
In the presence of a non-vanishing angular momentum $L$ of the $B_q$ 
decay products, this factor has properly to be taken into account 
in the determination of the sign of $\sin(\phi_q+\gamma)$ from 
$\langle S_q\rangle_+$.

\item Should the width difference $\Delta\Gamma_s$ be sizeable, the 
combination of the ``tagged'' mixing-induced observables 
$\langle S_s\rangle_\pm$ with their ``untagged''
counterparts $\langle {\cal A}_{\Delta\Gamma_s}\rangle_\pm$ offers an
elegant determination of $\tan(\phi_s+\gamma)$ in an essentially 
unambiguous manner, which does not require knowledge of $x_s$. Another 
important aspect of untagged rate measurements is the efficient 
determination of the hadronic parameters $x_q$. To accomplish this 
task, we may apply various untagged strategies, which do not rely 
on a sizeable value of $\Delta\Gamma_q$. 

\item We have derived bounds on $\phi_q+\gamma$, which can straightforwardly 
be obtained from the mixing-induced $B_q\to D_q\overline{u}_q$ observables, 
and provide essentially the same information as the ``conventional'' 
determination of $\phi_q+\gamma$, which suffers from multiple discrete
ambiguities. Giving a few examples, we have illustrated the potential 
power of these constraints, and have seen that stringent bounds on $\gamma$ 
may be obtained through a {\it combined} study of $B_s\to D_s\overline{u}_s$ 
and $B_d\to D_d\overline{u}_d$ modes. 

\item If we perform a simultaneous analysis of $U$-spin-related decays, for 
example of the $B_s\to D_s^{(\ast)\pm}K^\mp$, $B_d\to D^{(\ast)\pm}\pi^\mp$ 
system, we may follow various attractive avenues to determine $\gamma$ from 
the corresponding mixing-induced observables $\langle S_q\rangle_\pm$. The 
differences between these methods are due to different implementations of the 
$U$-spin relations for the hadronic parameters $a_q$ and $\delta_q$. For 
example, we may extract $\gamma$ by assuming $\tan\delta_s=\tan\delta_d$ or 
$a_s=a_d$. In comparison with the conventional $B_q\to D_q\overline{u}_q$ 
approaches, the most important advantage of these strategies -- apart from 
features related to discrete ambiguities -- is that $x_d$ does not have to 
be fixed, and that $x_s$ may only enter through $1+x_s^2$, i.e.\ a moderate 
correction, which can straightforwardly be included through untagged $B_s$ 
rate measurements; an accurate determination of $x_d$ and $x_s$ would only 
be interesting for the inclusion of $U$-spin-breaking corrections to 
$a_s/a_d$. After various steps of refinement, we would eventually arrive at 
an unambiguous, theoretically clean value of $\gamma$, and could also 
obtain -- as a by-product -- valuable insights into $U$-spin-breaking effects.
\end{itemize}
Since $B_{s,d}\to D_{s,d}\overline{u}_{s,d}$ modes will be accessible in 
the era of the LHC, in particular at LHCb, we strongly encourage a 
simultaneous analysis of $B_s$ and $B_d$ modes -- especially of 
$U$-spin-related decay pairs -- to fully exploit their very interesting 
physics potential.

\end{document}